\documentclass[%
 superscriptaddress,
 nofootinbib,
 amsmath,amssymb,
 aps,
 twocolumn, 
 prl,
 floatfix,
]{revtex4-2}

\usepackage{overpic}
\usepackage{float}
\usepackage{dcolumn}
\usepackage{bm}
\usepackage{braket}
\usepackage{xcolor}
\usepackage{soul}
\usepackage{subcaption}
\usepackage{mathtools}
\usepackage{MnSymbol}
\usepackage{bm}
\usepackage[utf8]{inputenc}
\usepackage[T1]{fontenc}
\usepackage{physics}
\usepackage{dsfont}
\usepackage{amsthm}
\usepackage{hyperref}
\usepackage{graphicx}
\usepackage{algorithm}
\usepackage{algcompatible}
\usepackage{algpseudocode}
\usepackage{mathrsfs}
\usepackage{amssymb}
\usepackage{qcircuit}
\usepackage{enumerate}

\theoremstyle{plain}

\theoremstyle{definition}

\theoremstyle{remark}

\makeatletter 
    
\renewcommand\onecolumngrid{
\do@columngrid{one}{\@ne}%
\def\set@footnotewidth{\onecolumngrid}
\def\footnoterule{\kern-6pt\hrule width 1.5in\kern6pt}%
}

\renewcommand\twocolumngrid{
        \def\footnoterule{
        \dimen@\skip\footins\divide\dimen@\thr@@
        \kern-\dimen@\hrule width.5in\kern\dimen@}
        \do@columngrid{mlt}{\tw@}
}%

\makeatother  

\begin{document}
\preprint{APS/123-QED}

\title{Optimal Strategies for Multi-parameter Quantum Metrology}

\author{Linxuan Li}
\email{lxli@link.cuhk.edu.hk}
\affiliation{Department of Mechanical and Automation Engineering, The Chinese University of Hong Kong, Shatin, Hong Kong SAR, China}
\author{Zihao Hu}
\affiliation{Department of Mechanical and Automation Engineering, The Chinese University of Hong Kong, Shatin, Hong Kong SAR, China}
\author{Longyun Chen}
\email{longyun@connect.hku.hk}
\affiliation{
 QICI Quantum Information and Computation Initiative, School of Computing and Data Science, The University of Hong Kong, Pokfulam Road, Hong Kong SAR, China
}
\author{Yuxiang Yang}
\email{yuxiang@cs.hku.hk}
\affiliation{
 QICI Quantum Information and Computation Initiative, School of Computing and Data Science, The University of Hong Kong, Pokfulam Road, Hong Kong SAR, China
}
\author{Haidong Yuan} 
\email{hdyuan@mae.cuhk.edu.hk}
\affiliation{Department of Mechanical and Automation Engineering, The Chinese University of Hong Kong, Shatin, Hong Kong SAR, China}
\affiliation{The Hong Kong Institute of Quantum Information Science and Technology, The Chinese University of Hong Kong, Shatin, Hong Kong SAR, China}
\affiliation{State Key Laboratory of Quantum Information Technologies and Materials, The Chinese University of Hong Kong, Shatin, Hong Kong SAR, China}

\date{\today}

\begin{abstract}
Estimating multiple unknown parameters simultaneously is essential for practical quantum sensing. However, it faces a fundamental challenge: the optimal strategy for estimating one parameter is often incompatible with that for another, making it impossible to simultaneously achieve the ultimate precision limits for all parameters. Here we develop a general and efficient computational framework that jointly optimizes probe states, control operations, and measurements across different strategy families, including parallel, sequential, and those with indefinite causal order. Our approach provides exact semidefinite-program formulations for several precision bounds, including the Holevo, Nagaoka–Hayashi, and quantum Cram\'er-Rao bounds. We demonstrate the capabilities of the framework in multiparameter magnetometry and frequency estimation, identifying optimal protocols within each class and revealing a strict hierarchy among the achievable performances of different classes in the multiparameter regime. The framework also directly incorporates resource constraints, such as energy budgets, enabling systematic investigation of experimentally realistic sensing scenarios. Furthermore, we develop a finite-memory optimization method for sequential strategies with restricted ancillary-memory dimension. By decomposing the protocol into initial probe preparation and intermediate control operations, this method provides a practical route to designing resource-constrained sequential sensing schemes. Our work establishes a versatile computational tool for determining fundamental precision limits and designing optimal quantum-sensing protocols in complex multiparameter settings.
\end{abstract}

\maketitle

\section{Introduction}
Quantum metrology \cite{helstrom1969quantum,braunstein1994statistical, holevo2011probabilistic} harnesses entanglement, squeezing, and exotic causal structures \cite{oreshkov2012quantum,chiribella2013quantum,araujo2015witnessing} to push measurement precision beyond classical limits, approaching the fundamental Heisenberg bound \cite{giovannetti2004quantum,giovannetti2011advances,degen2017quantum}. A typical metrological protocol involves three stages: probe preparation \cite{pezze2018quantum,yang2020probe,zheng2022preparation,le2023variational}, intermediate control \cite{PhysRevLett.115.110401,yuan2016sequential,Pang_2017,demkowicz2017adaptive,niu2019universal,valeri2020experimental}, and measurement \cite{braun2018quantum,yang2019optimal,zhou2023optimal,wang2025tight}. Their joint optimization is essential for achieving ultimate precision. Developing efficient  methods for this joint optimization is therefore of both fundamental and practical importance.

To capture all three stages—probe preparation, control, and measurement—within a single optimization, a unified description is essential. The framework of quantum strategies, often formulated in terms of quantum combs or testers, provides such a description. With quantum combs providing the necessary language, significant progresses have been made in identifying the optimal strategies for single-parameter quantum metrology \cite{liu2023optimal,liu2024fully,kurdzialek2403quantum,PhysRevResearch.6.L032048,Bavaresco_2024,43wz-mtrk,dulian2025qmetro++}. 
the success in the single-parameter domain, however, does not translate directly to the multi-parameter regime \cite{matsuzaki2011magnetic,dorfman2016nonlinear,chao2016fisher,moreau2019imaging,hou2020minimal, hou2021super,Zhang_2026}. There, a fundamental roadblock emerges: incompatibility \cite{chen2022incompatibility,albarelli2022probe,guhne2023colloquium,hu2024controlincompatibilitymultiparameterquantum,Chen_2024,wang2025tight}. No single protocol can achieve the highest precision for all parameters simultaneously. Instead, any realistic protocol must navigate a fundamental trade-off, where improving precision for one parameter inevitably comes at the expense of another. Consequently, the design of strategies becomes significantly more complex, as the trade-off between different parameters must be carefully balanced. To characterize the ultimate precision attainable in multi-parameter estimation for realistic scenarios, and quantify the incompatibility between the optimal strategies tailored to distinct parameters, a general and efficient computational framework for the optimization of multi-parameter strategies is therefore highly desirable.

In this work, we address this challenge by developing a suite of semi-definite programming (SDP) algorithms to compute various bounds, including the Holevo bound \cite{holevo2011probabilistic,Albarelli_2019,demkowicz2020multi}, the Nagaoka–Hayashi (NH) bound \cite{conlon2021efficient}, and the quantum Cram\'er Rao bound \cite{helstrom1969quantum,braunstein1994statistical, holevo2011probabilistic,hayashi2005asymptotic,liu2020quantum}. Extending the conic programming framework~\cite{hayashi2024finding} from the state estimation to channel estimation, our approach jointly optimizes probe states and control operations to minimize weighted estimation error. An important advantage of this formulation is that it not only provides a multiparameter extension of quantum-comb optimization, but also offers a natural way to study strategies under physically motivated resource constraints. While unrestricted ancillary memory and arbitrary controls are useful for identifying ultimate precision limits, they can be unrealistic in near-term sensing platforms, where the available memory dimension and control energy are limited. We therefore also consider resource-constrained formulations, including finite-memory sequential strategies in which the initial probe preparation and intermediate control operations are optimized in a decomposed manner under a fixed memory dimension. This makes it possible to compare the ideal precision limit with the performance attainable under finite-memory or control-resource restrictions, and to quantify the precision loss caused by experimentally relevant resource constraints. These results extends the hierarchy of quantum strategies to the multi-parameter regime and provides a practical tool for analyzing compatibility, verifying theoretical limits, and designing optimal sensing protocols.

\section{Preliminaries}
\label{sec_preliminary}
We consider the estimation of a vector of unknown parameters $\bm{\theta}=(\theta_1,\dots,\theta_d)^T$ encoded in a quantum channel $\mathcal{E}_{\bm{\theta}}$. The channel acts on an initial probe state, and measurements are performed on the output state. From the measurement outcomes $x$, one can construct an estimator $\hat{\bm{\theta}}(x)$. In this study, we focus on the locally unbiased estimators that satisfy
\begin{align} 
    &\sum_x \mathrm{Tr}[\rho_{\bm{\theta}} M_x] \hat{\theta}_i(x) = \theta_i,\label{lu1}\\
    &\frac{\partial}{\partial \theta_j} \sum_x \mathrm{Tr}[\rho_{\bm{\theta}} M_x] \hat{\theta}_i(x) = \delta_{ij},\label{lu2}
\end{align}
where $\{M_x|\sum_xM_x=I\}$ is the Positive-Operator-Valued-Measurement (POVM).
The precision of the estimation can be quantified by the mean-square error matrix $V$ with 
\begin{equation}
    V=\sum_x \mathrm{Tr}(\rho_{\bm{\theta}} M_x)(\hat{\bm{\theta}}(x)-\bm{\theta})(\hat{\bm{\theta}}(x)-\bm{\theta})^T.
\end{equation}
We note that the constraint in Eq. (\ref{lu1}) can actually be removed without loss of generality. To see this, consider any estimator $\hat{\bm{\theta}}$ that satisfies only the derivative condition Eq.(\ref{lu2}), we can define a recentered estimator $\tilde{\bm{\theta}}:=\hat{\bm{\theta}}-\bm{b}(\bm{\theta},\hat{\bm{\theta}})$, where $\bm{b}(\bm{\theta},\hat{\bm{\theta}}):=\sum_x p_{\bm{\theta}}(x)\hat{\bm{\theta}}-\bm{\theta}$ is the bias of the estimator. This recentering does not change the derivative condition. Under this transformation, the weighted mean-square error satisfies $\mathrm{Tr}[WV(\tilde{\bm{\theta}})]=\mathrm{Tr}[WV(\hat{\bm{\theta}})]-\bm{b}^TW\bm{b}\le \mathrm{Tr}[WV(\hat{\bm{\theta}})]$. Therefore, for $W\ge0$, removing Eq.(\ref{lu1}) does not affect the minimum achievable WMSE \cite{Sidhu_2020}.

The quantum Cramér–Rao bound (QCRB), also known as the symmetric-logarithmic-derivative (SLD) Cramér-Rao bound, provides a lower bound on the attainable $V$ as 
$V \ge \frac{1}{m} \mathcal{F}_{\bm{\theta}}^{-1}$, where $m$ is the number of repeated measurements and $\mathcal{F}_{\bm{\theta}}$ is the quantum Fisher information matrix (QFIM) \cite{helstrom1969quantum,holevo2011probabilistic}. In the single-parameter estimation, the QCRB is asymptotically saturable. For multiple parameters, however, the QCRB is generally not saturable. This is due to the incompatibility of the optimal measurements for estimating different parameters, which necessitates trade-offs among the achievable precisions for different parameters. To capture these trade-offs, one typically minimizes the weighted mean-square error (WMSE) $\mathrm{Tr}[W V]$, where $W$ is a $d\times d$ positive weight matrix. 

Several lower bounds have been proposed to characterize the attainable value of \(\operatorname{Tr}[WV]\). The most direct benchmark is the \emph{tight bound},
defined as the minimum value of the cost directly over all POVMs and locally unbiased
estimators \cite{hayashi2005asymptotic,hayashi2023tight}:
\begin{equation}
C^{\mathrm{tight}}
:=
\min_{\{M_x\},\,\hat{\bm\theta}}
\left\{
\operatorname{Tr}[WV(\hat{\bm{\theta}},M_x)]
\,\middle|\,
\text{(\ref{lu1}) and (\ref{lu2}) satisifed.}
\right\}.
\end{equation}
By definition, \(C^{\mathrm{tight}}\) is the lowest achievable weighted
error and is attainable in principle. However, it is usually difficult
to evaluate directly.

A more tractable lower bound is the \emph{Holevo bound} \cite{holevo2011probabilistic, nagaoka2005new, hayashi2023tight, demkowicz2020multi}, which is given by 
\begin{equation}
C^{\mathrm H}
:=
\min_{V,\bm X}
\left\{
\operatorname{Tr}[WV]
\,\middle|\,
V\ge Z[\bm X],\ 
\operatorname{Tr}\!\left[\nabla \rho_{\bm\theta}\bm X
\right]
=
\mathcal{I}
\right\},
\end{equation}
here \(V\) is a real symmetric matrix,
\(\bm X=(X_1,\ldots,X_d)^T\) is a vector of Hermitian operators and $Z[X]_{ij}=\operatorname{Tr}(\rho_{\bm\theta}X_iX_j)$. 
The Holevo bound is asymptotically saturable when collective measurements on arbitrary copies of the states can be performed under certain regularity conditions~\cite{Yamagata_2013,Yang_2019}. 

If only separable measurements can be performed, the Holevo bound is also not attainable in general. In this case, the
\emph{Nagaoka-Hayashi} \cite{nagaoka2005new,hayashi2023tight,conlon2021efficient} bound provides a tighter bound, which is given by 
\begin{equation}
C^{\mathrm{NH}}
:=
\min_{\bm L,\bm X}
\left\{
\operatorname{Tr}[(W\otimes\rho_{\bm\theta})\bm L]
\,\middle|\,
L_{jk}=L_{kj},\ 
\bm L\ge \bm X\bm X^T\right\},
\end{equation}
here \(\bm X=(X_1,\ldots,X_d)^T\) is a vector of
Hermitian operators that satisfy the locally unbiased condition $\partial_{\bm{\theta}_i}\text{Tr}[\rho_{\bm{\theta}}X_j]=\delta_{ij}$ for $i,j=1,\dots,d$, and \(\bm L\) is a $d\times d$ block matrix with each entry $L_{jk}$ a Hermitian operator that has the same dimension as $\rho_\theta$. These bounds satisfy the following hierarchy \cite{conlon2024gappersistencetheoremquantum}
\begin{equation}
C^{\mathrm{tight}}\ge C^{\mathrm{NH}}\ge C^{\mathrm H}\ge C^{\mathrm{SLD}}.
\end{equation}

\begin{figure}[t]

    \centering
    \captionsetup{justification=raggedright,singlelinecheck=false}

    {\label{subfig:parallel strategy}%
        \begin{overpic}[height=0.25\linewidth,keepaspectratio]{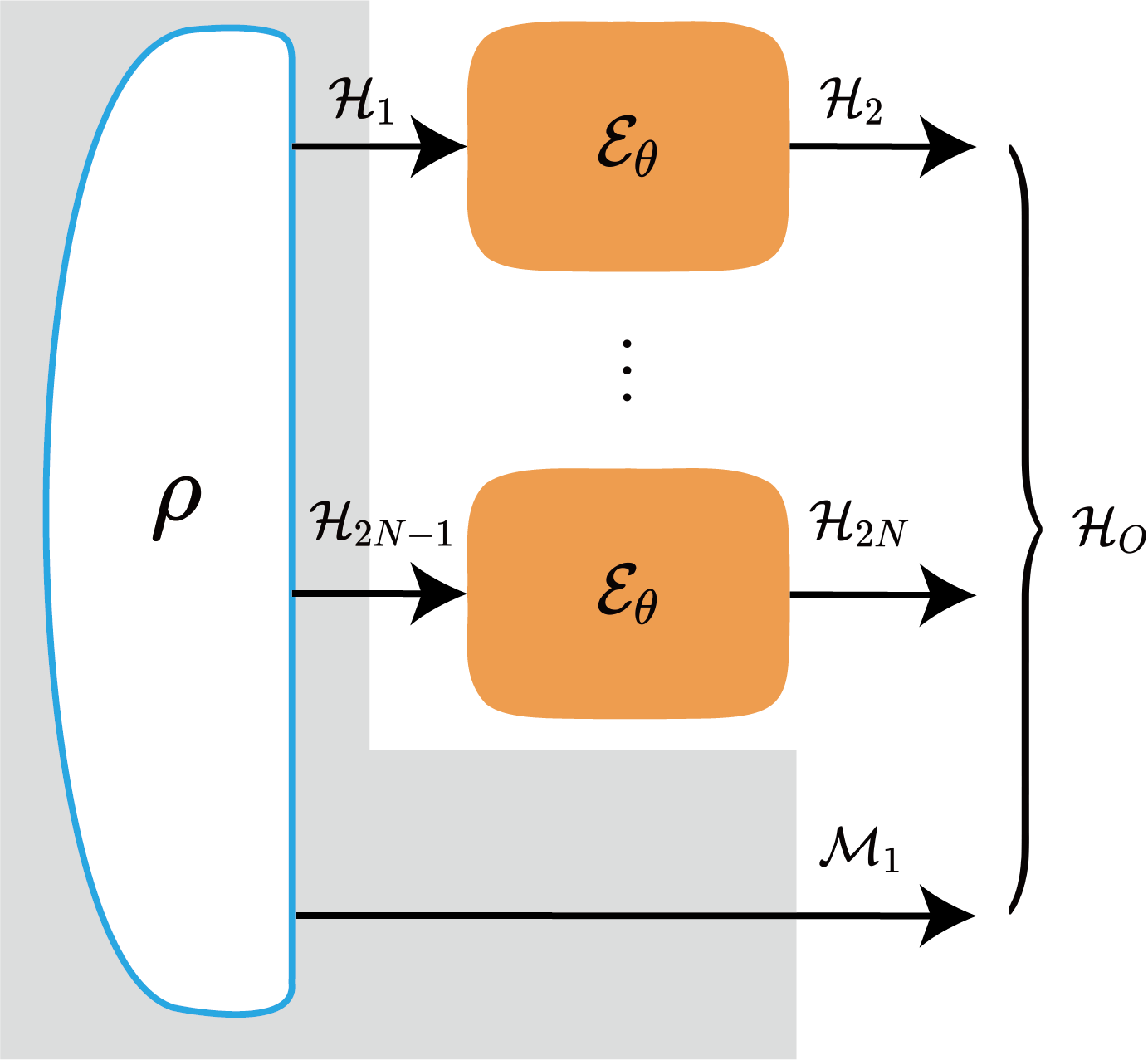}
            \put(-19,96){(a)}
        \end{overpic}}
    \hspace{0.02\linewidth}
    {\label{subfig:sequential strategy}%
        \begin{overpic}[height=0.25\linewidth,keepaspectratio]{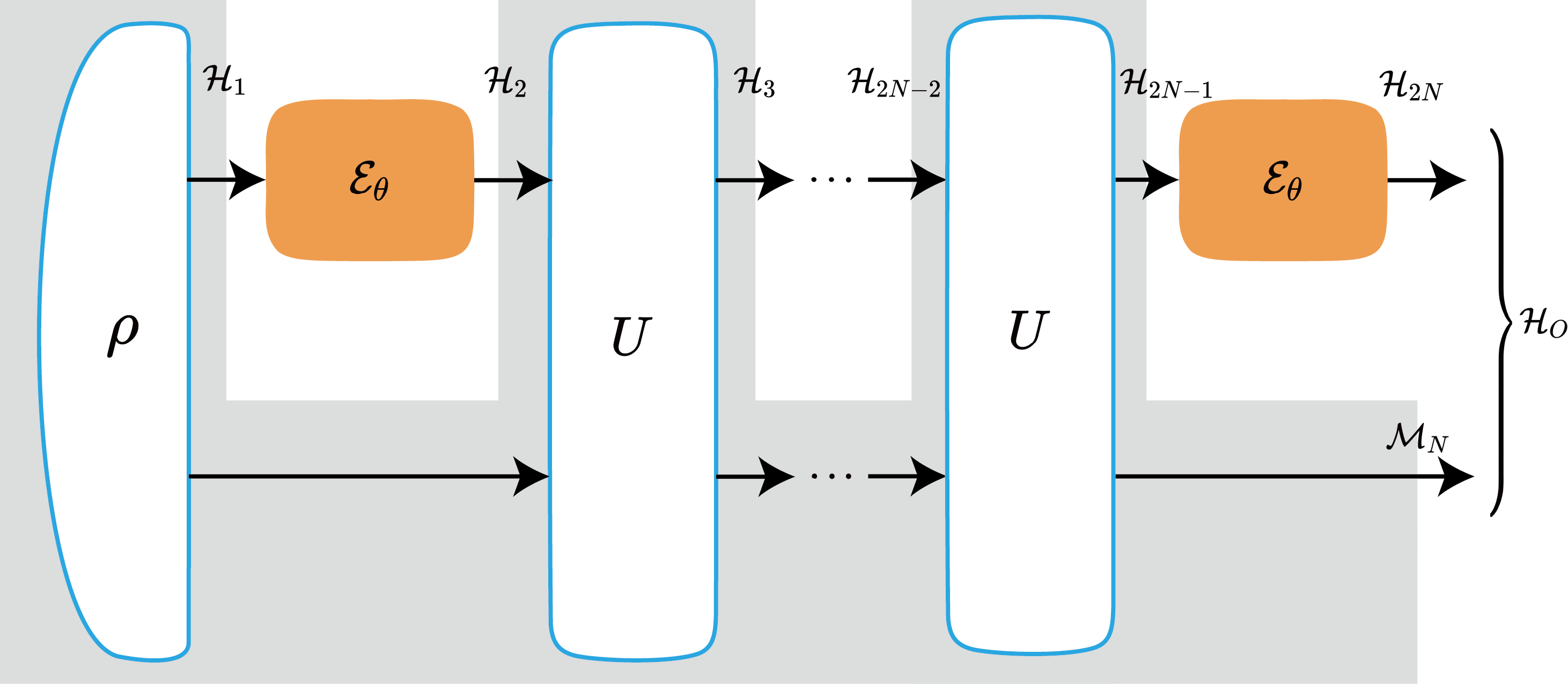}
            \put(-8,45){(b)}
        \end{overpic}}

    \vspace{0.5cm}
    {\label{subfig:quantum switch strategy}%
        \begin{overpic}[height=0.25\linewidth,keepaspectratio]{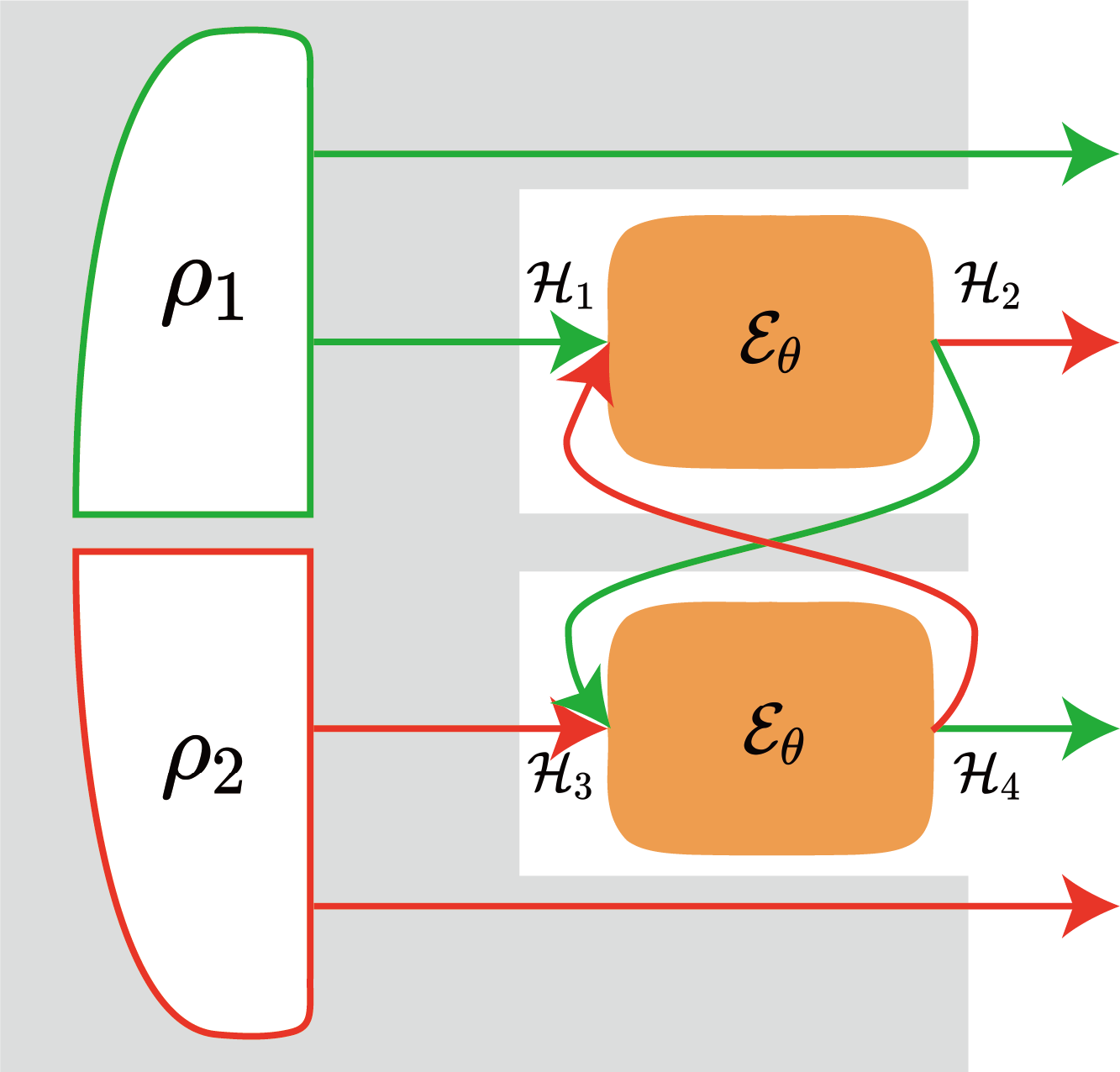}
            \put(-16,100){(c)}
        \end{overpic}}
    \hspace{0.02\linewidth}
    {\label{subfig:causal superposition strategy}%
        \begin{overpic}[height=0.25\linewidth,keepaspectratio]{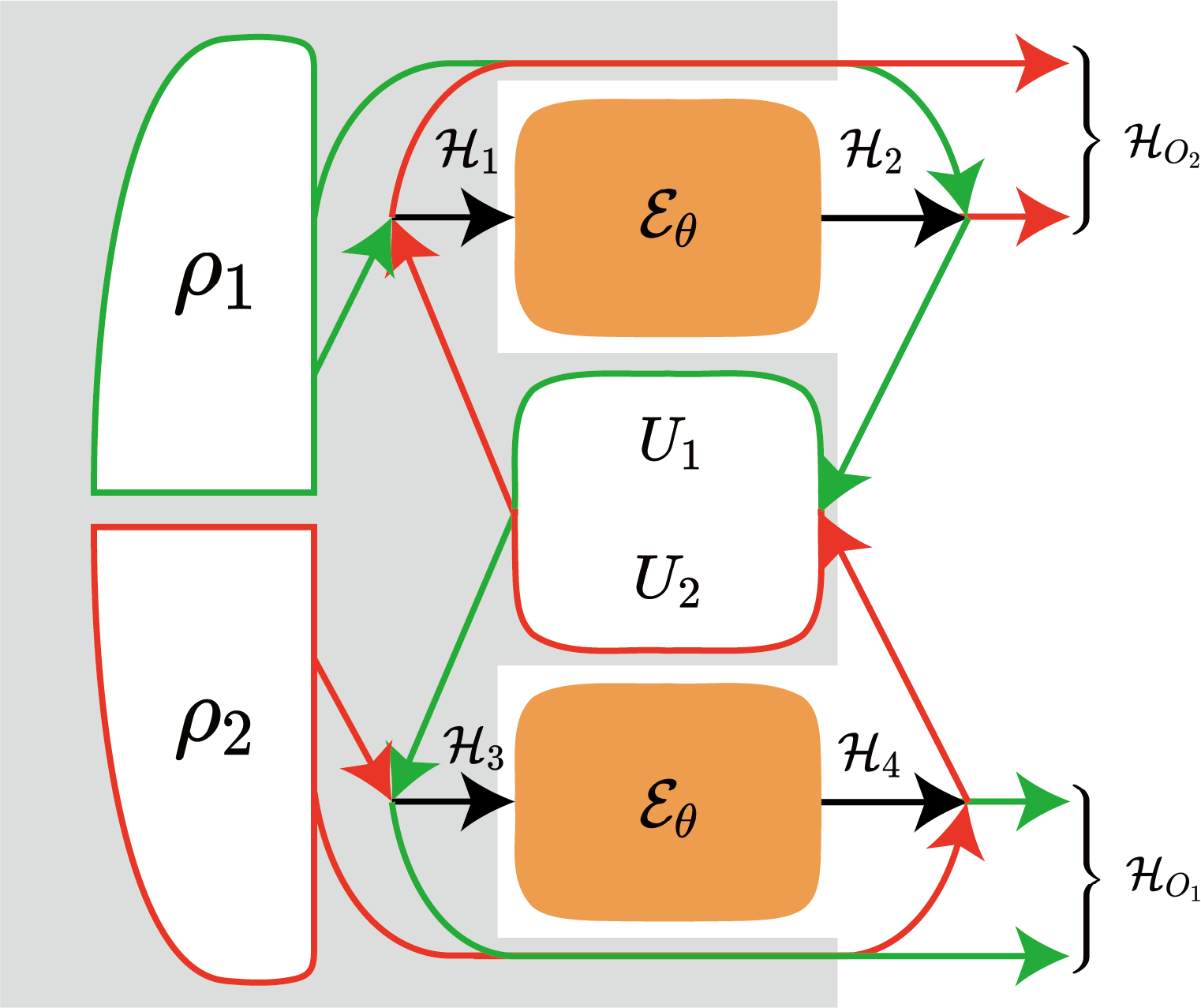}
            \put(-16,88){(d)}
        \end{overpic}}
    \caption{\textbf{Prototypical strategies of quantum metrology for $N$ channels}. $\mathcal E_{\bm{\theta}}$ is a quantum channel embedded unknown parameters $\bm{\theta}$, and the shaded grey area represents a quantum strategy $\tilde{P}$. (a) Parallel strategy (\texttt{Par}). (b) Sequential strategy (\texttt{Seq}), where $U_1,\dots,U_N$ are control operations. (c) Quantum SWITCH (\texttt{SWI}). The green and red lines, respectively, correspond to two different causal orders with no intermediate controls. (d) Causal superposition strategy (\texttt{Sup}), which is composed of two sequential strategies, plotted in green and in red respectively.}
    \label{fig_strategies}
\end{figure}

When multiple accesses to the channel are available, we need to consider different strategies to arrange the channels. This includes the parallel and sequential strategies, as shown in Fig.(\ref{fig_strategies}). The parallel strategy can be considered as a special case of the sequential strategy with the intermediate controls taken as the SWAP operations between the system and different ancillas. These strategies can be described using the quantum comb formalism.

For example, a general sequential strategy with $N$ uses of the channel can be described by 
an $N$-slot comb $\tilde{P}$:
\begin{equation}
\begin{aligned}
\tilde{P}\in 
\mathsf{Comb}\big[&
(\emptyset,\mathcal{H}_1),
(\mathcal{H}_2,\mathcal{H}_3),
\dots,\\
&(\mathcal{H}_{2N-2},\mathcal{H}_{2N-1}\otimes \mathcal{M}_N)\big],
\end{aligned}
\end{equation}
here each pair of Hilbert spaces specifies the input-output spaces of a tooth of the comb. The first tooth $(\emptyset,\mathcal{H}_1)$ corresponds to initial probe preparation, while intermediate teeth $(\mathcal{H}_{2j},\mathcal{H}_{2j+1})$ represent control operations that process the output of the $j$-th channel and prepare the input for the $(j+1)$-th channel. The final output space includes a memory system $\mathcal{M}_N$.
Different strategies are characterized by different constraints on $\tilde{P}$. For the sequential strategy, the comb needs to satisfy the following recursive relation:
\begin{equation}\label{eq:seqcomb}
    \begin{aligned}
        &\mathrm{Tr}_{\mathcal{M}_N}[\tilde{P}]=P^{(N)},\\
        &\mathrm{Tr}_{2k-1}[P^{(k)}]=\mathcal{I}_{2k-2}\otimes P^{(k-1)},~k=N,\dots,2,\\
        &\text{Tr}[P^{(1)}]=1.
    \end{aligned}
\end{equation}
The constraints for other strategies, including the parallel and the indefinite causal order(ICO) \cite{chiribella2013quantum,oreshkov2012quantum,araujo2015witnessing,Chapeau_Blondeau_2021,mukhopadhyay2018superpositioncausalordermetrological}, can be found in Appendix \ref{appendix:Expressions for Quantum Strategies}.

The concatenation of the quantum channels with the quantum comb can be described with the Choi-Jamiołkowski (CJ) operators using the link product \cite{chiribella2008quantum,chiribella2009theoretical}. Physically, the link product represents the operation of connecting the output space of one quantum operation to the input space of another one, while tracing over the connected Hilbert spaces. For two operators $A\in\mathcal{L}(\mathcal{H}_{\mathcal{A}})$ and $B\in\mathcal{L}(\mathcal{H}_{\mathcal{B}})$, the link product is defined as
\begin{equation}
A* B:=\mathrm{Tr}_{\mathcal{A} \cap \mathcal{B}}
\left[
\left(\mathcal{I}_{\mathcal{B}\setminus \mathcal{A}}\otimes 
A^{T_{\mathcal{A} \cap \mathcal{B}}}\right)
\left(B \otimes \mathcal{I}_{\mathcal{A}\setminus \mathcal{B}}\right)
\right],
\end{equation}
where $T_{\mathcal{A} \cap \mathcal{B}}$ denotes the partial transpose on the joint space 
$\mathcal{H}_{\mathcal{A}} \cap \mathcal{H}_{\mathcal{B}}$. When 
$\mathcal{H}_{\mathcal{A}} \cap \mathcal{H}_{\mathcal{B}}=\emptyset$, the link product reduces to the tensor product, namely $A*B=A\otimes B$. The final state with $N$ queries of the channel can then be obtained by the link product of the $N$-slot quantum comb with $N$ CJ operators of the channel as 
\begin{equation}
\rho_{\bm{\theta}}
=N_{\bm{\theta}}
*
\tilde{P},
\end{equation}
where 
\begin{equation}\label{eq:Ntheta}
N_{\bm{\theta}}=E^{(1)}_{\bm{\theta}}*
E^{(2)}_{\bm{\theta}}*
\dots *
E^{(N)}_{\bm{\theta}}
\end{equation}
with $E_\theta =
\mathcal{E}_\theta \otimes \mathcal{I}
\left(|I \rrangle \llangle I|\right)$ as the CJ operator of the channel $\mathcal{E}_{\bm{\theta}}$, here $|I \rrangle$ denotes the unnormalized maximally entangled state. 
For Markovian process where the $N$ uses of the channel acting on mutually distinct Hilbert spaces with 
$\mathcal{H}_j \cap \mathcal{H}_k = \emptyset$ for $j\neq k$, 
\(N_{\bm{\theta}}=\bigotimes_{i=1}^N E_{\bm{\theta}}^{(i)}\). We denote $\mathcal{H}_S:=\bigotimes_{i=1}^{2N}\mathcal{H}_{i}$ as the total input and output space for $N$ channels.

\section{Conic Optimization of Definite Causal Order Strategies}
\label{sec_conic}
For convenience, we denote the output space for final encoded state as
 \(   \mathcal{H}_O=\mathcal{H}_{O_S}\otimes \mathcal{H}_{O_M},\)
here $\mathcal{H}_{O_S}$ denotes the output system originating from the external channel, and $\mathcal{H}_{O_M}$ denotes the ancillary memory system retained by the strategy comb before measurement, as shown in Fig.~\ref{fig_strategies}. For the sequential strategy, $\mathcal{H}_{O_S}=\mathcal{H}_{2N}$ and $\mathcal{H}_{O_M}=\mathcal{M}_N$, while for the parallel strategy, $\mathcal{H}_{O_S}=\bigotimes_{i=1}^N\mathcal{H}_{2i}$ and $\mathcal{H}_{O_M}=\mathcal{M}_1$.

To identify the optimal strategy, we first recast the WMSE in a form suitable for conic optimization over quantum strategies. The cost function is given by
\begin{equation}
    \mathrm{Tr}[WV]
    =
    \sum_x 
    \mathrm{Tr}\!\left[
        W\rho_{\bm{\theta}}
        \bigl(\hat{\bm\theta}(x)-\bm\theta\bigr)
        \bigl(\hat{\bm\theta}(x)-\bm\theta\bigr)^T
        M_x
    \right],
\end{equation}
where $\rho_\theta=N_\theta*\tilde{P}$ is the output state, \(\hat{\boldsymbol{\theta}}(x)\) is the locally unbiased estimator and \(\{M_x\}\) is the POVM.  Notably, the estimator and the measurement operators appear jointly in this expression. To combine them into a single optimization variable, we introduce the auxiliary operator
\begin{equation}
    X :=
    \sum_x
    \begin{bmatrix}
        1 & \Delta\bm{\theta}(x)^T \\
        \Delta\bm{\theta}(x) &
        \Delta\bm{\theta}(x)\Delta\bm{\theta}(x)^T
    \end{bmatrix}
    \otimes M_x
    \in \mathcal{L}(H_\mathcal{R}\otimes \mathcal{H}_O),
\end{equation}
where $\Delta\bm{\theta}(x)=\hat{\bm{\theta}}(x)-\bm{\theta}$, $H_\mathcal{R}\simeq\mathbb{R}^{d+1}$ is a real reference space of dimension \(d+1\). With this construction, the minimization of the WMSE can be expressed equivalently as
\begin{equation}\label{eq:QP}
    \min_{\tilde{P}\in \mathcal{\tilde{P}},X}
    \mathrm{Tr}\!\left[
        \bar{W}\otimes (N_{\bm{\theta}}*\tilde{P})X
    \right],
\end{equation}
where $\mathcal{\tilde{P}}:=\{\tilde{P}\big | \tilde{P}\ge 0, P:=\mathrm{Tr}_{O_m}[\tilde{P}]\in \texttt{seq/par}\}$ 
is the set of certain strategies with the corresponding constraints defined explicitly in Appendix \ref{appendix:Expressions for Quantum Strategies}, $\bar{W}=[0]\oplus W$ is a block-diagonal matrix. The proof of the equivalence between the original WMSE and this new expression is provided in Appendix~\ref{appendix:Proof for Q}. The normalization of the POVM $\sum_x M_x=I$ becomes a linear constraint on $X$ as
\begin{equation}
    \mathrm{Tr}_{\mathcal{R}}
    \left[
        (|0\rangle \langle 0|_{\mathcal{R}} \otimes \mathcal{I}_O) X
    \right]
    =
    \mathcal{I}_O,
    \label{Q:locally-unbiased i}
\end{equation}
while the local-unbiasedness condition in Eq.~(\ref{lu2}) becomes
\begin{equation}
    \frac{1}{2}
    \mathrm{Tr}
    \left[
        \left(
            (|0\rangle \langle i|_{\mathcal{R}}
            +
            |i\rangle \langle 0|_{\mathcal{R}})
            \otimes
            (\partial_{\theta_j}N_{\bm{\theta}}*\tilde{P})
        \right)X
    \right]
    =
    \delta_{ij},
    \label{Q:locally-unbiased ii}
\end{equation}
for $i,j=1,\ldots,d$. Thus, the problem is formulated as
\begin{equation}
    Q^{P}
    :=
    \min
    \mathrm{Tr}\!\left[
        \bar{W}\otimes (N_{\bm{\theta}}*\tilde{P})X
    \right],
\end{equation}

Optimizing $X$ over different cones $\mathcal{C}_k$, $k=1,2,3,4$, yields the tight, NH, SLD and Holevo bounds, respectively \cite{hayashi2023tight,hayashi2024finding}. The explicit definitions of these cones and their connections to the corresponding bounds can be found in Appendix~\ref{appendix:Expressions for Cones}. For instance, the tight bound is obtained by optimizing $X$ over the separable cone
\begin{equation}
   \mathcal{C}_1
    :=
    \mathrm{conv}\left\{W\otimes Z \bigg| W\in \mathcal{L}(\mathcal{R})_{rs,+}, Z\in \mathcal{B}(\mathcal{H}_O)_+\right\},
\end{equation}
where $\mathcal{L}(\mathcal{R})_{rs,+}$ is the set of real symmetric positive semidefinite matrices on $\mathcal{R}$, and $\mathcal{B}(\mathcal{H}_O)_+$ is the set of bounded complex positive semidefinite operators on $\mathcal{H}_O$. 

Due to the bilinear terms involving $X$ and $\tilde{P}$ in the objective function of Eq.(\ref{eq:QP}), the original problem is still not convex. For strategies with definite causal order, the bilinear term can be eliminated by introducing a new variable that combines $X$ and $\tilde{P}$,
\begin{equation}
\label{defination_omega}
    \Omega
    :=
    \mathrm{Tr}_{O_M}
    \left[
        \left(
            \tilde{P}
            \otimes
            \mathcal{I}_{\mathcal{R},O_S}
        \right)
        \left(
            \mathcal{I}_{\bar{S}}
            \otimes
            X
        \right)
    \right]\in \mathcal{L}(\mathcal{H_R}\otimes \mathcal{H}_S),
\end{equation}
where $\mathcal{H}_{\bar{S}}:=\mathcal{H}_{S}/\mathcal{H}_{O_{\mathrm{S}}}$.

The normalization and l.u. constraints can now be expressed directly in terms of $\Omega$ with
\begin{eqnarray}\label{S:locally-unbiased i}
    (\mathrm{i}):&\mathrm{Tr}_{\mathcal{R}}
    \left[
        \Omega
        \left(
            |0\rangle \langle 0|_{\mathcal{R}}
            \otimes
            \mathcal{I}_{S}
        \right)
    \right]
    =
    P\otimes I_{O_S},\\
\label{S:locally-unbiased ii}
    (\mathrm{ii}):&\frac{1}{2}
    \mathrm{Tr}
    \left[
        \Omega
        \left(
            |0\rangle \langle i|_{\mathcal{R}}
            +
            |i\rangle \langle 0|_{\mathcal{R}}
        \right)
        \otimes
        \frac{\partial T_{\bm{\theta}}}{\partial \theta^j}
    \right]
    =
    \delta_{ij},
\end{eqnarray}
for $i,j=1,\ldots,d$, here
 \(   T_{\bm{\theta}}
    =
    N_{\bm{\theta}}^{\mathrm{T}_{\bar{S}}},\)
with $\mathrm{T}_{\bar{S}}$ denoting the partial transpose on $\mathcal{H}_{\bar{S}}$. 

Importantly, the cone constraint is preserved under this construction. To see this more explicitly, we note that the construction in
Eq.~(\ref{defination_omega}) defines a linear map $\Phi_{\tilde{P}}(\cdot):=\mathrm{Tr}_{O_M}[\sqrt{\tilde{P}
            \otimes
            \mathcal{I}_{O_S}}(\cdot)\sqrt{\tilde{P}
            \otimes
            \mathcal{I}_{O_S}}]:
    \mathcal{L}
    \left(
        \mathcal{H}_{O}
    \right)
    \to
    \mathcal{L}
    \left(
        \mathcal{H}_{S}
    \right)$ such that
\begin{equation}
    \Omega
    =
    \left(
        \mathcal{I}_{\mathcal{R}}
        \otimes
        \Phi_{\tilde{P}}
    \right)X.
\end{equation}
We first consider the tight-bound cone. If
$X\in\mathcal{C}_{1}$, then $X$ admits a separable decomposition
\begin{equation}
    X
    =
    \sum_{\mu}
    W_\mu
    \otimes
    Z_\mu,
    \qquad
    W_\mu\in\mathcal{L}(\mathcal{R})_{rs,+},
    \quad
    Z_\mu\in\mathcal{B}(\mathcal{H}_{O})_+ .
\end{equation}
Under the strategy contraction, we obtain
\begin{equation}
    \Omega
    =
    \left(
        \mathcal{I}_{\mathcal{R}}
        \otimes
        \Phi_{\tilde{P}}
    \right)(X)
    =
    \sum_{\mu}
    W_\mu
    \otimes
    \Phi_{\tilde{P}}(Z_\mu).
\end{equation}
Since $\Phi_{\tilde{P}}$ is a positive map, each positive operator
$Z_\mu\in\mathcal{B}(\mathcal{H}_{O})_+$ is mapped to a positive operator
$\Phi_{\tilde{P}}(Z_\mu)\in\mathcal{B}(\mathcal{H}_{S})_+$. For element in the separable cone, it is sufficient that every term in the above decomposition remains a product of positive operators. Hence 
\begin{equation}
    \Omega
    \in
    \mathrm{conv}
    \left\{
        W\otimes Z
        ~\middle|~
        W\in\mathcal{L}(\mathcal{R})_{rs,+},
        \;
        Z\in \mathcal{B}(\mathcal{H}_{S})_+
    \right\}.
\end{equation}
This is a cone in
$\mathcal{H}_\mathcal{R}\otimes \mathcal{H}_{S}$, which we denote by $\tilde{\mathcal{C}}_{1}$. For the remaining cones corresponding to the NH, SLD and Holevo bound, we can similarly obtain the cones $\tilde{\mathcal{C}}_{k}$ such that 
\begin{equation} X \in \mathcal{C}_{k} ~~\Rightarrow~~ \Omega = \left( \mathcal{I}_{\mathcal{R}} \otimes \Phi_{\tilde{P}} \right)(X) \in \tilde{\mathcal{C}}_{k}, \end{equation}
for $k=2,3,4$. Conversely, every feasible $\Omega\in \tilde {\mathcal{C}}_k$ can be obtained from a feasible pair of $(P,X)$ with $X\in \mathcal{C}_k$. Detailed proof can be found in Appendix \ref{appendix:Proof of S_k^P<=Q_k^P}. 

With the new variable, we can reformulate the minimization of WMSE in Eq.(\ref{eq:QP}) as
\begin{equation}
\label{theorem2-obj}
    S_k^{P}
    :=
    \min_{\substack{
        P\in\mathcal{P},~\Omega\in \tilde{\mathcal{C}}_k 
    }}
        \mathrm{Tr}
        \left[
            \Omega
            (W\otimes T_{\bm{\theta}})
        \right]
\end{equation}
where $\mathcal{P}:=\{P\big|P\ge 0, P\in \texttt{seq/par}\}$ and $\Omega$ satisfies conditions (i) and (ii) given in Eq.(\ref{S:locally-unbiased i}) and (\ref{S:locally-unbiased ii}). In Appendix~\ref{appendix:proof of theorem 1}, we provide explicit proof that
\begin{equation}
    S_k^{P}=Q_k^{P},
\end{equation}
where $Q_k^{P}$ corresponds to the tight, NH, SLD and Holevo bound for $k=1,2,3,4$ respectively. This provides an equivalent conic formulation where the objective function is now a linear function of the new variable $\Omega$.

Once an optimal solution $\Omega^\star$ is obtained, the corresponding strategy operator can be reconstructed from condition (i) as
\begin{equation}
\label{opt_P}
    P^\star
    =
    \frac{1}{d_{O_S}}
    \mathrm{Tr}_{\mathcal{R},O_S}
    \left[
        \Omega^*
        \left(
            \ket{0}\bra{0}
            \otimes
            \mathcal{I}_{S}
        \right)
    \right].
\end{equation}
A physical realization of the optimal strategy can then be obtained from $P^\star$ by decomposing it into quantum circuits \cite{bisio2011minimal,PhysRevA.93.032318, iten2021introductionuniversalqcompiler}. We note that the conditions for $\mathcal{C}_k$, $k=2,3,4$--which correspond to NH, SLD, and Holevo bound--can be formulated as linear constraints and therefore be efficiently calculated with semi-definite programming (SDP). In contrast, the separable cone $\mathcal{C}_1$, can not be efficiently characterized with finite linear constraints, which is related to the computational hardness of characterizing separable states, in the other word, charactering the entangled states. 
However, efficient algorithms with SDP for the upper and lower bounds of the tight bound have been developed in previous study for the state estimation \cite{hayashi2023tight}. 

For a general indefinite-causal-order (ICO) strategy, the normalization condition in \(S_k^P\) can be formulated as
\begin{equation}
\mathrm{Tr}_{\mathcal{R}}
\left[
\Omega
\left(
|0\rangle \langle 0|_{\mathcal{R}}
\otimes
\mathcal{I}_{S}
\right)
\right]
=
P_{\mathrm{ICO}},
\end{equation}
where \(P_{\mathrm{ICO}}\in\mathcal{L}(\mathcal{H}_S)\) obeys the process matrix formalism ~\cite{oreshkov2012quantum}, as introduced in Appendix \ref{appendix:Expressions for Quantum Strategies}. Hence, the general ICO strategy can be computed within the same conic optimization framework. However, the physical implementation of a general ICO is untraceable~\cite{Wechs_2021,PhysRevLett.127.110402}. We therefore consider physically motivated subclasses such as causal superposition~\cite{araujo2014computational,wechs2021quantum,kurdzialek2023using} and the quantum SWITCH~\cite{PhysRevA.88.022318,rubino2017experimental}. These physically realizable ICO subclasses can be expressed in terms of the superposition of definite-causal-order strategies. For example, for a causal superposition strategy, one may write
\begin{equation}
P_{\mathrm{ICO}}
=
\sum_\pi
q_\pi
P_\pi
\otimes
\mathcal{I}_{2\pi(N)},
\end{equation}
where \(\pi\) denotes a permutation, \(q_\pi\ge0\), \(\sum_\pi q_\pi=1\), and \(P_\pi\) is the definite causal order strategy associated with \(\pi\). Directly optimizing over both \(q_\pi\) and \(P_\pi\) would introduce a non-convex coupling between the weights and the corresponding definite-order strategies. To avoid the issue, we absorb the weights into the strategy operators by defining
\begin{equation}
P^q_{\pi}:=q_\pi P_\pi .
\end{equation}
Then condition~\(\mathrm{(i)}\) becomes the affine constraint
\begin{equation}
\mathrm{Tr}_{\mathcal{R}}
\left[
\Omega
\left(
|0\rangle \langle 0|_{\mathcal{R}}
\otimes
\mathcal{I}_{S}
\right)
\right]
=
\sum_\pi
P^q_{\pi}
\otimes
\mathcal{I}_{2\pi(N)} .
\end{equation}
For each \(\pi\), \(P^q_\pi\) obeys the homogeneous definite causal order constraints. For example, when \(P^q_\pi\) corresponds to a sequential strategy, it satisfies
\begin{equation}
\begin{aligned}
\mathrm{Tr}_{2\pi(N)-1}
\left[
P_\pi^{q,(N-1)}
\right]
&=
\mathcal{I}_{2\pi(N)-2}
\otimes
P_\pi^{q,(N-2)},\\
&\cdots\\
\mathrm{Tr}
\left[
P_\pi^{q,(1)}
\right]
&=
q_\pi,
\end{aligned}
\end{equation}
where
\begin{equation}
\sum_\pi q_\pi=1 .
\end{equation}
This provides a systematic procedure to calculate the multi-parameter bounds for ICO strategies.

We consider a multi-parameter quantum magnetometry problem. The dynamics is generated by the Hamiltonian $H=B\bm{n}\cdot \bm{\sigma}$, where $B$ is the field strength, $\bm{n} = (\sin\theta\cos\phi,\sin\theta\sin\phi,\cos\theta)$ specifies the field direction, and $\bm{\sigma}=(\sigma_x,\sigma_y,\sigma_z)$ denotes the Pauli operators. The encoding unitary takes the form $U_{\bm{\theta}} = e^{-i\alpha \bm{n}\cdot \bm{\sigma}}$ with $\alpha = Bt$. Here $(\alpha,\theta,\phi)$ are the unknown parameters to  be estimated. 

To go beyond the ideal unitary setting, we also include amplitude-damping (AD) noise:
\begin{equation}\label{eq:AD}
\mathcal{E}^{\text{(AD)}}_{\bm{\theta}}(\rho)= K_0 U_{\bm{\theta}}\rho U_{\bm{\theta}}^\dagger K_0^\dagger+K_1 U_{\bm{\theta}}\rho U_{\bm{\theta}}^\dagger K_1^\dagger ,
\end{equation}
where $K_0= \ket{0}\bra{0} + \sqrt{1-p}\ket{1}\bra{1}$, $K_1 = \sqrt{p}\ket{0}\bra{1}$.
This provides a simple yet informative setting for comparing parallel, sequential, and indefinite-causal-order strategies in a genuine multi-parameter estimation task. 

\begin{figure}[t]
    \centering
    \captionsetup{justification=raggedright,singlelinecheck=false}

    \begin{subfigure}[b]{0.48\linewidth}
        \begin{overpic}[width=\linewidth]{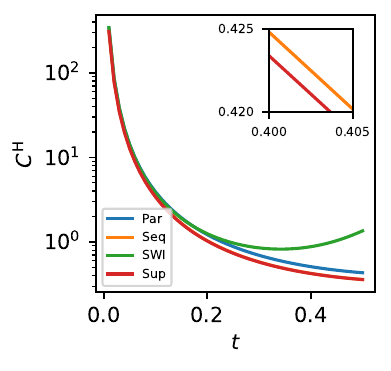}
            \put(3,92){(a)}
        \end{overpic}
        \label{fig1:a}
    \end{subfigure}
    \hfill
    \begin{subfigure}[b]{0.49\linewidth}
        \begin{overpic}[width=\linewidth]{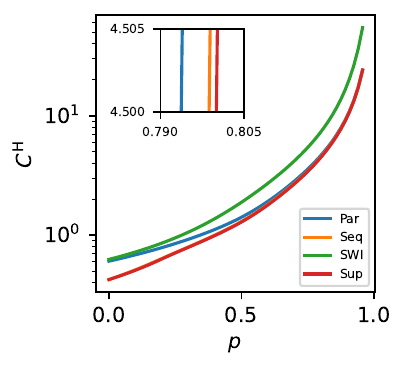}
            \put(3,92){(b)}
        \end{overpic}
        \label{fig1:b}
    \end{subfigure}

    \vspace{0.5em}

    \begin{subfigure}[b]{0.98\linewidth}
        \begin{overpic}[width=\linewidth]{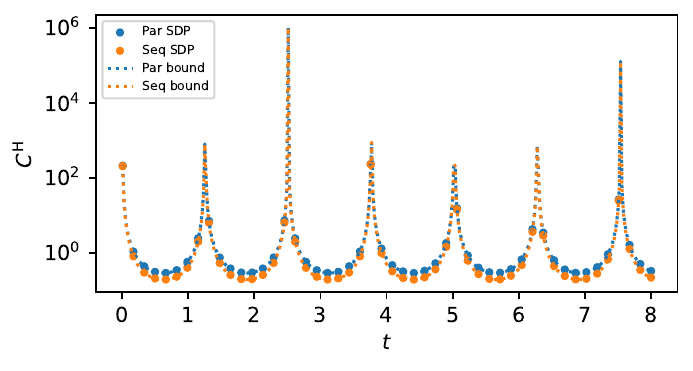}
            \put(1.2,55){(c)}
        \end{overpic}
        \label{fig1:c}
    \end{subfigure}

    \caption{\textbf{Holevo bound using parallel \texttt{Par}, sequential \texttt{Seq}, and indefinite-causal-order strategies \texttt{Sup} for causal superposition and \texttt{SWI} for quantum SWITCH.} We take $N = 2$, $B=2.5$, $\theta=1.3$, and $\phi = \pi/4$. (a): fix the decay noise rate $p=0.25$ and vary the evolution time $t$. (b): fix $t=1.0$ and vary the decay noise rate $p$. (c): comparison between the numerical and analytical results of Eqs.~(\ref{ana_par}) and (\ref{ana_adp}) in the noiseless case, where the circle markers ``SDP'' represent the numerical results and the dashed lines ``bound'' represent the analytical results.}
    \label{fig1:all}
\end{figure}

We focus on the computation of the Holevo bound since it is asymptotically attainable under collective measurements. Fig.~\ref{fig1:all} summarizes the results for \(N=2\) and \(W=I\). In the noisy regime Fig.~1(a) and 1(b), the strategies exhibit a stable performance ordering. The gap between causal superposition and general ICO is negligible under a tolerance of $10^{-7}$ so we only show the curve for the causal superposition strategy (\texttt{Sup}). Causal superposition and sequential strategies perform nearly identically and both outperform the remaining strategies, yielding the hierarchy $C^{\texttt{Sup}}<C^{\texttt{Seq}}<C^{\texttt{Par}}<C^{\texttt{SWI}}$. Besides, $C^{\texttt{Sup}}\simeq C^{\texttt{Seq}}<C^{\texttt{Par}}\simeq C^{\texttt{SWI}}$ when $p\to 0$ and $C^{\texttt{Sup}}\simeq C^{\texttt{Seq}}\simeq C^{\texttt{Par}}< C^{\texttt{SWI}}$ when $p\to 1$ for this particular setting of Fig.~1(b).

In the noiseless case, the Holevo bound, the NH bound and the tight bound all coincide. 
Moreover, the optimal performances of the parallel and sequential schemes are known analytically, providing a useful benchmark for our numerical results. For ancilla-assisted parallel strategies, the minimal WMSE is given analytically by \cite{hou2020minimal}
\begin{equation}
\label{ana_par}
w_\alpha \delta\hat{\alpha}^2+w_\theta\delta\hat{\theta}^2+w_\phi \delta\hat{\phi}^2
\ge
\frac{\left(\sqrt{w_\alpha}+\frac{\sqrt{w_\theta}}{|\sin\alpha|}+\frac{\sqrt{w_\phi}}{|\sin\alpha \sin\theta|}\right)^2}{4N(N+2)}.
\end{equation}
For sequential strategies, the minimal WMSE is \cite{yuan2016sequential}
\begin{equation}
\label{ana_adp}
w_\alpha \delta\hat{\alpha}^2+w_\theta\delta\hat{\theta}^2+w_\phi \delta\hat{\phi}^2
\ge
\frac{w_\alpha+w_\theta\sin^2\alpha+w_\phi\sin^2\alpha \sin^2\theta}{4N^2}.
\end{equation}
As shown in Fig. \ref{fig1:all}(c), our numerical results agree with these analytical expressions for both protocols. Furthermore, the numerically obtained optimal probe state and optimal control also agree with the analytical results obtained in previous studies \cite{yuan2016sequential,hou2020minimal}. The numerical method thus correctly recovers the known optimal bounds in the analytically solvable regime. Detailed discussions are provided in Appendix~\ref{appendix:NV for seq and par}.

\section{Resources-constrained Strategies}
The optimal strategies presented in the previous section were derived without imposing any resource restrictions. In practical implementations, however, resource constraints are unavoidable. We now extend our method to accommodate such constraints, thereby providing a framework for designing resource‑efficient optimal strategies.

Let $\mathfrak{M}$ denote a resource measure on the strategy set. The constraint on the admissible strategies under a given budget $E\ge 0$, can be written as 
\begin{equation}
\mathcal{P}_E := \left\{ P\in\mathcal{P}\middle| \mathfrak{M}(P) \leq E \right\}.
\end{equation}
Since the optimization variable $\Omega$ is connected to the strategy operator $P$ via a linear map in Eq.~\eqref{S:locally-unbiased i}. Consequently, any linear resource constraints on $P$ translates directly into linear constraints on $\Omega$.
Specifically, when the constraint $\mathfrak{M}(P) \leq E$ is linear, it can be incorporated directly as linear constraints on $\Omega$ as
\begin{equation}
    \mathfrak{M}\left(
    \frac{1}{d_{O_S}}\mathrm{Tr}_{\mathcal{R},O_S}
    \left[
        \Omega
        \left(
            \ket{0}\bra{0}
            \otimes
            \mathcal{I}_{S}
        \right)
    \right] \right) = \mathfrak{M}(P) \leq E,
\end{equation}
and directly incorporated into the conic programming.

As a concrete illustration, we take $\mathfrak{M}$ to be the energy consumed by the initial state preparation and the intermediate control operations included in the strategy, quantified within the global battery model introduced in Ref.~\cite{Chen_2026}. In this model, a single battery system, which can be implemented by a sufficiently large-dimensional quantum system, supplies the energy consumed in all steps of the strategy, including initial state preparation and intermediate controls. Each step is implemented through a joint energy-preserving operation involving the battery system. Therefore, by energy conservation, the energy consumed by the strategy can be characterized by the energy lost by the battery system, which equals the increase in the energy of the outputs relative to the inputs during the steps.
In our setting, each input and output qubit of the parameterized channels is assigned the Hamiltonian $\ket{1}\bra{1}$ with the ground-state energy set to zero, while all ancillary systems are assumed to be degenerate. Consequently, arbitrary operations on the ancillas are energy-free, and energy costs are incurred only by the qubits in the channel space $\mathcal{H}_S$.

Following the characterization in Ref.~\cite{Chen_2026}, a strategy is feasible in the global battery model if the cumulative energy increase from inputs to outputs over the first $n$ steps never exceeds the available budget for $n=1,\dots,N$. When the energy budget is $E \geq 0$, for a sequential strategy $P = P^{(N)}$, this condition is equivalent to
\begin{equation}
\mathfrak{M}(P) := \max_{n=1,2,\dots,N} \lambda_{\max}\left( O_{n} \right) \leq E
\end{equation}
with $O_{n} := \operatorname{Tr}_{1,3,\dots,2n-1}
\left[ P^{(n)} \left(\mathcal{I}_{2,4,\dots,2n-2}\otimes H_{1,3,\dots,2n-1}\right) \right] - \left(H_{2,4,\dots,2n-2}\right)^T$ and $\lambda_{\max}(O_n)$ being the largest eigenvalue of $O_n$. Here, $H_{1,3,\dots,2n-1}$ and $H_{2,4,\dots,2n-2}$ denote the total Hamiltonians of the spaces $\bigotimes_{i=1}^{n} \mathcal{H}_{2i-1}$ and $\bigotimes_{i=1}^{n} \mathcal{H}_{2i-2}$, respectively. 
For a parallel strategy $P$, since no intermediate control is inserted between the parameterized channels, the energy constraint reduces to bounding the initial state energy as
\begin{equation}
\mathfrak{M}(P) := \operatorname{Tr}[H_{1,3,\dots,2N-1} P] \leq E.
\end{equation}
For a causal superposition strategy $P=\{q_{\pi} P_{\pi}\}_{\pi}$, one may consider the spacetime-individual global battery model in Ref.~\cite{Chen_2026}. In this model, each branch of the causal order is required to be feasible with its own battery of budget $E$, namely
\begin{equation}
\mathfrak{M}(P) := \max_{\pi} \mathfrak{M}(P_{\pi})\le E.
\end{equation}

\begin{figure}[t!]
    \centering
    \includegraphics[width=\linewidth]{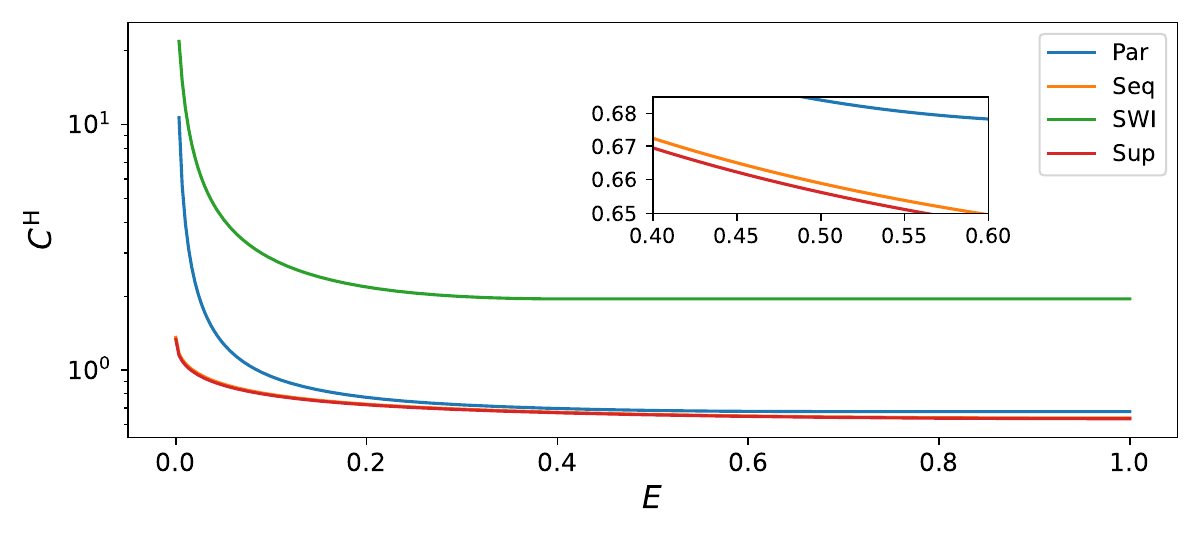}
    \caption{\textbf{Energy-constrained Holevo bound using parallel \texttt{Par}, sequential \texttt{Seq}, and indefinite-causal-order strategies \texttt{Sup} for causal superposition and \texttt{SWI} for quantum SWITCH.} 
    We fix $N=2$, $p=0.5$, $\theta=1.3$, $\phi=\pi/4$, $B=2.5$, and $t=0.6$ (hence $\alpha=1.5$), while varying the energy budget $E \geq 0$.
}
    \label{fig:bound-vs-energy}
\end{figure}

We illustrate our approach with the multi-dimensional quantum magnetometry problem of Sec.~\ref{sec_conic}, imposing an explicit energy budget $E$ on the admissible strategies. Fig.~\ref{fig:bound-vs-energy} plots the Holevo bound versus $E$, showing that the bound decreases monotonically with increasing energy—confirming energy as an operational metrological resource. The figure also shows a clear hierarchy among strategy classes, indicating that causal superposition retains its energy-efficiency advantage in the multi-parameter setting, consistent with the single-parameter result \cite{Chen_2026}.

\section{Optimal strategy with finite memory}

Although conic programming provides the optimal strategy, its practical applicability is limited by the rapid growth in the number of variables with the number of channel uses, restricting direct numerical implementation to small \(N\). A natural simplification is to restrict the strategy to a finite‑dimensional memory—a choice that is also physically justified, as ancillary memory systems are scarce experimental resources. With a fixed memory dimension, the problem reduces to optimizing the initial probe preparation and intermediate control combs, offering a computationally tractable finite‑memory approximation to the full sequential optimization.

Under the finite‑memory assumption, we set the memory spaces to have fixed dimension: \(\mathcal{M}_k \simeq \mathcal{M}\) and \(\dim(\mathcal{M}) = d_M\) for all \(1 \le k \le N\). This restriction allows us to decompose the global strategy \(\tilde{P}\) into a sequence of combs \(B_1, B_2, \dots, B_N\):
\[
\tilde{P} = B_1 \star B_2 \star \cdots \star B_N,
\]
where \(B_1 \in \mathcal{L}(\mathcal{H}_1 \otimes \mathcal{M}_1)\) represents the initial probe preparation, and \(B_k \in \mathcal{L}(\mathcal{H}_{2k-2} \otimes \mathcal{M}_{k-1} \otimes \mathcal{H}_{2k-1} \otimes \mathcal{M}_k)\) for \(2 \le k \le N\) represents the intermediate control at step \(k\). The problem can now be formulated as:
\begin{equation}
\begin{aligned}
    & Q^P_k=\min_{\{B_k\}_{k=1}^N,X}
    \mathrm{Tr}\!\left[
        \bar{W}\otimes \rho_{\bm{\theta}}X
    \right],\\
    s.t.
    &~ \mathrm{Tr}_{\mathcal{R}} \left[(|0\rangle \langle 0|_{\mathcal{R}} \otimes \mathcal{I}_O) X \right] = \mathcal{I}_O,\\
    & \frac{1}{2}\mathrm{Tr} \left[ \left(  (|0\rangle \langle i|_{\mathcal{R}} + |i\rangle \langle 0|_{\mathcal{R}}) \otimes \partial_{\theta_j}\rho_{\bm{\theta}} \right) X \right] = \delta_{i,j},\\
     &\rho_{\bm{\theta}}=B_1\star B_2\star \cdots \star B_N \star N_{\bm{\theta}},\\
     & B_1,B_k \ge 0,~\mathrm{Tr}(B_1)=1,\mathrm{Tr}_{\mathrm{out}_k}(B_k)=\mathcal{I}_{\mathrm{in}_k}, k=2, \cdots, N.
\end{aligned}
\end{equation}
where $1\le i,j\le d$ and $2\le k \le N$, $\mathrm{in}_k:=\mathcal{H}_{2k-2}\otimes \mathcal{M}_{k-1}$ and $\mathrm{out}_k:=\mathcal{H}_{2k-1}\otimes \mathcal{M}_{k}$ denote the input and output space of $B_k$ with $\mathsf{dim}(\mathcal{M}_k)=d_M$. The set of $\{B_k\}$ can be optimized iteratively. Similar iterative optimization methods have been adopted in previous work for single parameter estimation tasks \cite{liu2024efficient,kurdzialek2025quantum}, where the QFI can be treated as a multi-convex optimization over the combs $\{B_k\}_{k=1}^N$. 

The optimization problem above allows each control comb \(B_k\) to vary independently across steps. In many practical scenarios, however, quantum sensors have limited programmability, and it is often more efficient—or experimentally more feasible-to apply the same control operation at each step, i.e., \(B_k = B\) for all \(k = 2, \dots, N\). This identical‑control constraint significantly reduces the number of optimization variables, but it also introduces a nonlinearity: the contraction now involves \(N-1\) copies of the same comb \(B\), making the problem nonlinear in \(B\).

To retain tractable subproblems, we adopt a one-site surrogate
update strategy. At each iteration \(s\), a single control position
\(k_s\) is randomly selected for optimization, while all other
occurrences are frozen at the current control \(B^{(s)}\). The
resulting optimized candidate \(B_{\mathrm{cand}}\) defines the
interpolation path
\begin{equation}
    B(\lambda_s)
    =
    \cos^2(\pi\lambda_s)B^{(s)}
    +
    \sin^2(\pi\lambda_s)B_{\mathrm{cand}},
    ~
    \lambda_s\in\left[0,\frac{1}{2}\right].
\end{equation}
We then perform a line search along this path and select
\begin{equation}
    \lambda_s^\star
    =
    \operatorname*{arg\,min}_{0\leq\lambda_s\leq 1/2}
    Q^P_k\bigl(B(\lambda_s)\bigr).
\end{equation}
The control is subsequently updated as
\begin{equation}
\begin{aligned}
    B^{(s+1)}
    &=B(\lambda_s^\star)\\
    &=\cos^2(\pi\lambda_s^\star)B^{(s)}
    +
    \sin^2(\pi\lambda_s^\star)B_{\mathrm{cand}}.
\end{aligned}
\end{equation}
this ensures a controlled update and guarantees a
non-increasing objective value.

\begin{figure}[t]
    \centering
    \includegraphics[width=1\linewidth]{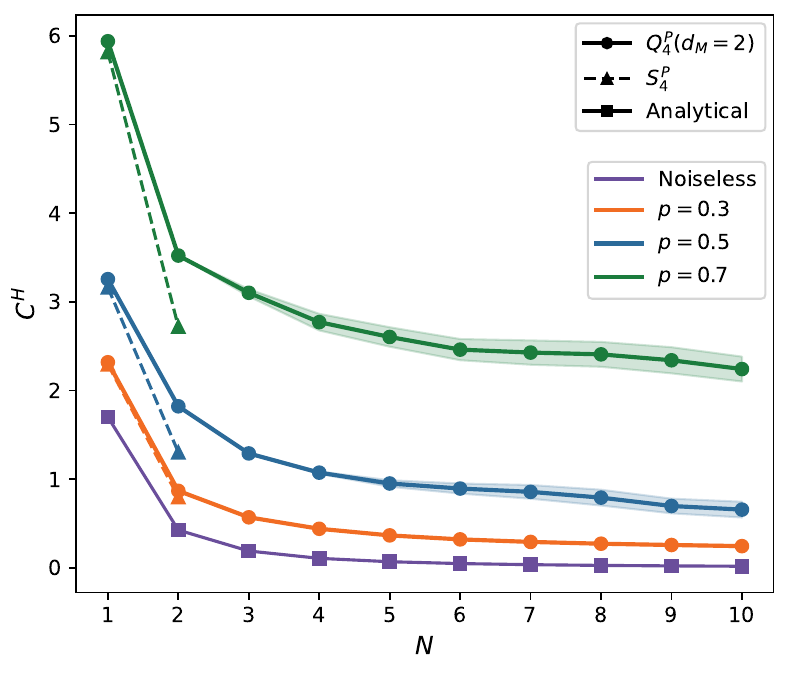}
    \caption{\textbf{Holevo bound $Q_k^P(d_M=2)$ versus $S^P_k$ considering optimization of finite-memory strategies with identical controls.} We consider 3-dimension quantum magnetometry problem in Sec.~\ref{sec_conic} and fix $\alpha=2.5$, $\theta=1.3$, and $\phi = \pi/4$. The analytical result indicates Eq.(\ref{ana_adp}) for noiseless case. Numerical simulations for $Q_k^P(d_M=2)$ are performed 20 times to compute the mean absolute errors and the corresponding 95\% confidence intervals, represented by solid lines and shaded regions.}
    \label{fig:identical_control_opt}
\end{figure}

The detailed procedure is given in Algorithm 1. Fig.~\ref{fig:identical_control_opt} displays the numerical results for the three‑dimensional magnetic field model from Sec.~\ref{sec_conic} under different levels of amplitude damping noise. At low noise, the identical‑control finite‑memory strategy performs nearly as well as the optimal‑control benchmark, showing that a fixed‑memory device with repeated controls can achieve most of the available precision. With increasing noise, however, the gap between the two strategies grows. We further note that the iterative optimization tends to become unstable in the high‑noise regime and may occasionally fail to converge, reflecting the increased complexity of the nonconvex optimization landscape.
 
\begin{algorithm}[H]
\caption{Optimization of finite-memory strategies with identical controls.}
\label{alg:identical-control}
\begin{enumerate}[{(i)}]
    \item Initialize a feasible set $\{B_1^{(0)},B^{(0)}\}$ and $d_M$.

    \item Given $\{B_1^{(s)},B^{(s)}\}$, compute $\rho^{(s)}$ and 
    $\{\partial_j\rho^{(s)}\}_{j=1}^q$, and solve the $X$-SDP to obtain 
    $X^{(s)}$.

    \item With $X^{(s)}$ and $B^{(s)}$ fixed, update $B_1^{(s)}$ by solving 
    the local $B_1$-SDP.

    \item Randomly choose one control position $k_s\in\{2,\ldots,N\}$. 
    With all other controls frozen to $B^{(s)}$, solve the one-site 
    $B_{k_s}$-SDP to obtain a candidate $B_{\mathrm{cand}}$.

    \item Perform minimization over $\lambda$ to update $B^{(s)}$ to $B^{(s+1)}$:
    \begin{equation*}
        B^{(s+1)}
        =B(\lambda_s^*)=
        \cos^2(\pi\lambda_s^*) B^{(s)}
        +
        \sin^2(\pi\lambda_s^*) B_{\mathrm{cand}},
    \end{equation*}
    where $\lambda_s^*=\operatorname*{arg\,min}_{0\leq\lambda_s\leq 1/2}
    Q^P_k\bigl(B(\lambda_s)\bigr)$ .
    
    \item Repeat (ii)--(v) until convergence. Return
    \begin{equation*}
        \left\{
        B_1^{(\mathsf{opt})},
        B^{(\mathsf{opt})},
        X^{(\mathsf{opt})},
        (Q_k^P)^{(\mathsf{opt})}
        \right\}.
    \end{equation*}
    
\end{enumerate}
\end{algorithm}

\section{Conclusion}

We have developed a systematic framework for optimizing multi-parameter quantum metrology over general classes of quantum strategies. For strategies with definite causal order, the ultimate precision bounds can be formulated as conic optimization problems, with different feasible cones reproducing the tight, Nagaoka--Hayashi, Holevo, and SLD bounds. For indefinite-causal-order (ICO) strategies, the framework applies to a broad class of protocols that admit decompositions into sequential or parallel strategy sectors, thereby extending the range of tractable optimization beyond fixed causal order.

We benchmarked the method by applying it to multi-dimensional quantum magnetometry. In the noiseless regime, the numerically optimized strategies saturate known analytical limits \cite{hou2020minimal,yuan2016sequential}, providing a direct validation of the formulation. In the presence of noise, the method reveals clear hierarchies among parallel, sequential, and indefinite-causal-order strategies, and identifies regimes in which causal superposition provides a genuine metrological advantage. The optimized probes and controls also match analytical predictions.

We further show that the approach can be adapted to resource‑constrained scenarios by restricting the ancillary memory to finite dimension and incorporating resource budgets directly into the optimization. This yields a tractable finite‑memory approximation that remains close to the optimal‑control benchmark in relevant parameter regimes, demonstrating the practical utility of the framework for experimentally realistic settings.

Despite its strengths, the approach has limitations. The dimension scales rapidly with both the number of channel uses $N$ and the number of parameters $d$, making large‑scale problems computationally demanding. To address this, we introduced a scheme with identical controls under a finite memory constraint, which has reduced complexity. While this scheme enhances scalability for multi-copy channels, it does so at the cost of reduced accuracy due to the loss of global optimality.

These limitations naturally motivate several avenues for future investigation. One promising direction is the extension of the current conic framework to related problems, such as quantum channel discrimination \cite {Sieniawski_2026} and Bayesian quantum metrology \cite {Bavaresco_2024}, which exhibit analogous strategy-optimization structures. A complementary direction is the development of more efficient numerical algorithms, potentially through the exploitation of problem-specific structure or the application of advanced convex optimization techniques, to accommodate larger-scale instances beyond the reach of current methods.

\textit{Note added.}---Upon conclusion of our work, we became aware of Ref.~\cite{zhou2026unifiedcomputableapproachoptimal}, which also studies multiparameter quantum metrology using the quantum tester formalism and SDP-computable conic methods, and Ref.~\cite{pang2026} which focus specifically on ICO strategies. The technical overlap with our work lies mainly in the use of conic optimization to compare different strategy classes. Ref.~\cite{zhou2026unifiedcomputableapproachoptimal} focuses on the tight Cramér--Rao type bound and computes SDP upper and lower bounds through separable-cone approximations, whereas our work develops exact SDP formulations for complementary precision bounds, including the Holevo, Nagaoka--Hayashi, and SLD bounds. Our framework further addresses resource constraints and practical implementability by incorporating energy budgets and developing a finite-memory optimization method for sequential strategies, motivated by realistic sensing platforms where ancillary memory and control resources are limited.

\section*{Acknowledgments}

We thank Qiushi Liu and Francesco Albarelli for insightful discussions. The numerical results are obtained via the Python packages CVXPY \cite{diamond2016cvxpy, agrawal2018rewriting} and MOSEK \cite{mosek} for convex optimization. H.Y. acknowledge the partial support from the Quantum Science and Technology-National Science and Technology Major Project (2023ZD0300600), the Research Grants Council of Hong Kong (14309223, 14309624, 14309022), Guangdong Provincial Quantum Science Strategic Initiative (GDZX2303007, GDZX2505003) and 1+1+1 CUHK-CUHK(SZ)-GDST Joint Collaboration Fund (Grant No. GRDP2025-022).
LC and YY acknowledge the support of the National Natural Science Foundation of China via the Excellent Young Scientists Fund (Hong Kong and Macau) Project 12322516, the National Natural Science Foundation of China (NSFC)/Research Grants Council (RGC) Joint Research Scheme via Project N\_HKU7107/24, and the Hong Kong Research Grant Council (RGC) via General Research Fund (GRF) grant 17302724.


\bibliography{apssamp}

@article{giovannetti2011advances,
  title={Advances in quantum metrology},
  author={Giovannetti, Vittorio and Lloyd, Seth and Maccone, Lorenzo},
  journal={Nature photonics},
  volume={5},
  number={4},
  pages={222--229},
  year={2011},
  publisher={Nature Publishing Group UK London}
}

@article{PhysRevA.88.022318,
  title = {Quantum computations without definite causal structure},
  author = {Chiribella, Giulio and D'Ariano, Giacomo Mauro and Perinotti, Paolo and Valiron, Benoit},
  journal = {Phys. Rev. A},
  volume = {88},
  issue = {2},
  pages = {022318},
  numpages = {15},
  year = {2013},
  month = {Aug},
  publisher = {American Physical Society},
  doi = {10.1103/PhysRevA.88.022318},
  url = {https://link.aps.org/doi/10.1103/PhysRevA.88.022318}
}

@article{oreshkov2012quantum,
  title={Quantum correlations with no causal order},
  author={Oreshkov, Ognyan and Costa, Fabio and Brukner, {\v{C}}aslav},
  journal={Nature communications},
  volume={3},
  number={1},
  pages={1092},
  year={2012},
  publisher={Nature Publishing Group UK London}
}

@article{degen2017quantum,
  title={Quantum sensing},
  author={Degen, Christian L and Reinhard, Friedemann and Cappellaro, Paola},
  journal={Reviews of modern physics},
  volume={89},
  number={3},
  pages={035002},
  year={2017},
  publisher={APS}
}

@article{giovannetti2004quantum,
  title={Quantum-enhanced measurements: beating the standard quantum limit},
  author={Giovannetti, Vittorio and Lloyd, Seth and Maccone, Lorenzo},
  journal={Science},
  volume={306},
  number={5700},
  pages={1330--1336},
  year={2004},
  publisher={American Association for the Advancement of Science}
}

@article{liu2023optimal,
  title={Optimal strategies of quantum metrology with a strict hierarchy},
  author={Liu, Qiushi and Hu, Zihao and Yuan, Haidong and Yang, Yuxiang},
  journal={Physical Review Letters},
  volume={130},
  number={7},
  pages={070803},
  year={2023},
  publisher={APS}
}

@article{liu2024fully,
  title={Fully-optimized quantum metrology: framework, tools, and applications},
  author={Liu, Qiushi and Hu, Zihao and Yuan, Haidong and Yang, Yuxiang},
  journal={Advanced Quantum Technologies},
  volume={7},
  number={12},
  pages={2400094},
  year={2024},
  publisher={Wiley Online Library}
}

@article{PhysRevResearch.6.L032048,
  title = {Strict hierarchy of optimal strategies for global estimations: Linking global estimations with local ones},
  author = {Zhou, Zhao-Yi and Qiu, Jing-Tao and Zhang, Da-Jian},
  journal = {Phys. Rev. Res.},
  volume = {6},
  issue = {3},
  pages = {L032048},
  numpages = {6},
  year = {2024},
  month = {Aug},
  publisher = {American Physical Society},
  doi = {10.1103/PhysRevResearch.6.L032048},

}

@article{kurdzialek2403quantum,
  title={Quantum metrology using quantum combs and tensor network formalism (2024)},
  author={Kurdzialek, Stanislaw and Dulian, Piotr and Majsak, Joanna and Chakraborty, Sagnik and Demkowicz-Dobrzanski, Rafal},
  journal={arXiv preprint arXiv:2403.04854}
}

@book{holevo2011probabilistic,
  title={Probabilistic and statistical aspects of quantum theory},
  author={Holevo, Alexander S},
  volume={1},
  year={2011},
  publisher={Springer Science \& Business Media}
}

@article{demkowicz2020multi,
  title={Multi-parameter estimation beyond quantum Fisher information},
  author={Demkowicz-Dobrza{\'n}ski, Rafa{\l} and G{\'o}recki, Wojciech and Gu{\c{t}}{\u{a}}, M{\u{a}}d{\u{a}}lin},
  journal={Journal of Physics A: Mathematical and Theoretical},
  volume={53},
  number={36},
  pages={363001},
  year={2020},
  publisher={IOP Publishing}
}

@article{liu2020quantum,
  title={Quantum Fisher information matrix and multiparameter estimation},
  author={Liu, Jing and Yuan, Haidong and Lu, Xiao-Ming and Wang, Xiaoguang},
  journal={Journal of Physics A: Mathematical and Theoretical},
  volume={53},
  number={2},
  pages={023001},
  year={2020},
  publisher={IOP Publishing}
}

@article{hayashi2023tight,
  title={Tight Cram{\'e}r-Rao type bounds for multiparameter quantum metrology through conic programming},
  author={Hayashi, Masahito and Ouyang, Yingkai},
  journal={Quantum},
  volume={7},
  pages={1094},
  year={2023},
  publisher={Verein zur F{\"o}rderung des Open Access Publizierens in den Quantenwissenschaften}
}

@article{conlon2021efficient,
  title={Efficient computation of the Nagaoka--Hayashi bound for multiparameter estimation with separable measurements},
  author={Conlon, Lorc{\'a}n O and Suzuki, Jun and Lam, Ping Koy and Assad, Syed M},
  journal={npj Quantum Information},
  volume={7},
  number={1},
  pages={110},
  year={2021},
  publisher={Nature Publishing Group UK London}
}

@article{chen2022incompatibility,
  title={Incompatibility measures in multiparameter quantum estimation under hierarchical quantum measurements},
  author={Chen, Hongzhen and Chen, Yu and Yuan, Haidong},
  journal={Physical Review A},
  volume={105},
  number={6},
  pages={062442},
  year={2022},
  publisher={APS}
}

@article{braunstein1994statistical,
  title={Statistical distance and the geometry of quantum states},
  author={Braunstein, Samuel L and Caves, Carlton M},
  journal={Physical Review Letters},
  volume={72},
  number={22},
  pages={3439},
  year={1994},
  publisher={APS}
}

@article{helstrom1969quantum,
  title={Quantum detection and estimation theory},
  author={Helstrom, Carl W},
  journal={Journal of statistical physics},
  volume={1},
  number={2},
  pages={231--252},
  year={1969},
  publisher={Springer}
}

@article{araujo2015witnessing,
  title={Witnessing causal nonseparability},
  author={Ara{\'u}jo, Mateus and Branciard, Cyril and Costa, Fabio and Feix, Adrien and Giarmatzi, Christina and Brukner, {\v{C}}aslav},
  journal={New Journal of Physics},
  volume={17},
  number={10},
  pages={102001},
  year={2015},
  publisher={IOP Publishing}
}

@manual{mosek,
   author = "MOSEK ApS",
   title = "The MOSEK Python Fusion API manual. Version 11.0.",
   year = 2025,
   url = "https://docs.mosek.com/latest/pythonfusion/index.html"
 }

@article{PhysRevLett.115.110401,
  title = {Optimal Feedback Scheme and Universal Time Scaling for Hamiltonian Parameter Estimation},
  author = {Yuan, Haidong and Fung, Chi-Hang Fred},
  journal = {Phys. Rev. Lett.},
  volume = {115},
  issue = {11},
  pages = {110401},
  numpages = {7},
  year = {2015},
  month = {Sep},
  publisher = {American Physical Society},
  doi = {10.1103/PhysRevLett.115.110401},
  url = {https://link.aps.org/doi/10.1103/PhysRevLett.115.110401}
}

@misc{mukhopadhyay2018superpositioncausalordermetrological,
      title={Superposition of causal order as a metrological resource for quantum thermometry}, 
      author={Chiranjib Mukhopadhyay and Manish K. Gupta and Arun Kumar Pati},
      year={2018},
      eprint={1812.07508},
      archivePrefix={arXiv},
      primaryClass={quant-ph},
      url={https://arxiv.org/abs/1812.07508}, 
}

@article{Chapeau_Blondeau_2021,
   title={Noisy quantum metrology with the assistance of indefinite causal order},
   volume={103},
   ISSN={2469-9934},
   url={http://dx.doi.org/10.1103/PhysRevA.103.032615},
   DOI={10.1103/physreva.103.032615},
   number={3},
   journal={Physical Review A},
   publisher={American Physical Society (APS)},
   author={Chapeau-Blondeau, François},
   year={2021},
   month=mar }

@article{Pang_2017,
   title={Optimal adaptive control for quantum metrology with time-dependent Hamiltonians},
   volume={8},
   ISSN={2041-1723},
   url={http://dx.doi.org/10.1038/ncomms14695},
   DOI={10.1038/ncomms14695},
   number={1},
   journal={Nature Communications},
   publisher={Springer Science and Business Media LLC},
   author={Pang, Shengshi and Jordan, Andrew N.},
   year={2017},
   month=mar }

@article{Wechs_2021,
   title={Quantum Circuits with Classical Versus Quantum Control of Causal Order},
   volume={2},
   ISSN={2691-3399},
   url={http://dx.doi.org/10.1103/PRXQuantum.2.030335},
   DOI={10.1103/prxquantum.2.030335},
   number={3},
   journal={PRX Quantum},
   publisher={American Physical Society (APS)},
   author={Wechs, Julian and Dourdent, Hippolyte and Abbott, Alastair A. and Branciard, Cyril},
   year={2021},
   month=aug }

@article{zhou2023optimal,
  title={Optimal protocols for quantum metrology with noisy measurements},
  author={Zhou, Sisi and Michalakis, Spyridon and Gefen, Tuvia},
  journal={PRX Quantum},
  volume={4},
  number={4},
  pages={040305},
  year={2023},
  publisher={APS}
}

@book{hayashi2005asymptotic,
  title={Asymptotic theory of quantum statistical inference: selected papers},
  author={Hayashi, Masahito},
  year={2005},
  publisher={World Scientific}
}

@misc{conlon2024gappersistencetheoremquantum,
      title={The gap persistence theorem for quantum multiparameter estimation}, 
      author={Lorcán O. Conlon and Jun Suzuki and Ping Koy Lam and Syed M. Assad},
      year={2024},
      eprint={2208.07386},
      archivePrefix={arXiv},
      primaryClass={quant-ph},
      url={https://arxiv.org/abs/2208.07386}, 
}

@incollection{nagaoka2005new,
  title={A new approach to Cram{\'e}r-Rao bounds for quantum state estimation},
  author={Nagaoka, Hiroshi},
  booktitle={Asymptotic theory of quantum statistical inference: Selected Papers},
  pages={100--112},
  year={2005},
  publisher={World Scientific}
}

@misc{iten2021introductionuniversalqcompiler,
      title={Introduction to UniversalQCompiler}, 
      author={Raban Iten and Oliver Reardon-Smith and Emanuel Malvetti and Luca Mondada and Gabrielle Pauvert and Ethan Redmond and Ravjot Singh Kohli and Roger Colbeck},
      year={2021},
      eprint={1904.01072},
      archivePrefix={arXiv},
      primaryClass={quant-ph},
      url={https://arxiv.org/abs/1904.01072}, 
}

@article{PhysRevA.93.032318,
  title = {Quantum circuits for isometries},
  author = {Iten, Raban and Colbeck, Roger and Kukuljan, Ivan and Home, Jonathan and Christandl, Matthias},
  journal = {Phys. Rev. A},
  volume = {93},
  issue = {3},
  pages = {032318},
  numpages = {19},
  year = {2016},
  month = {Mar},
  publisher = {American Physical Society},
  doi = {10.1103/PhysRevA.93.032318},
  url = {https://link.aps.org/doi/10.1103/PhysRevA.93.032318}
}

@article{kurdzialek2025quantum,
  title={Quantum metrology using quantum combs and tensor network formalism},
  author={Kurdzia{\l}ek, Stanis{\l}aw and Dulian, Piotr and Majsak, Joanna and Chakraborty, Sagnik and Demkowicz-Dobrza{\'n}ski, Rafa{\l}},
  journal={New Journal of Physics},
  volume={27},
  number={1},
  pages={013019},
  year={2025},
  publisher={IOP Publishing}
}

@article{liu2024efficient,
  title={Efficient tensor networks for control-enhanced quantum metrology},
  author={Liu, Qiushi and Yang, Yuxiang},
  journal={Quantum},
  volume={8},
  pages={1571},
  year={2024},
  publisher={Verein zur F{\"o}rderung des Open Access Publizierens in den Quantenwissenschaften}
}

@article{agrawal2018rewriting,
  author  = {Agrawal, Akshay and Verschueren, Robin and Diamond, Steven and Boyd, Stephen},
  title   = {A rewriting system for convex optimization problems},
  journal = {Journal of Control and Decision},
  year    = {2018},
  volume  = {5},
  number  = {1},
  pages   = {42--60},
}

@article{diamond2016cvxpy,
  author  = {Steven Diamond and Stephen Boyd},
  title   = {{CVXPY}: {A} {P}ython-embedded modeling language for convex optimization},
  journal = {Journal of Machine Learning Research},
  year    = {2016},
  volume  = {17},
  number  = {83},
  pages   = {1--5},
}

@article{Sidhu_2020,
   title={Geometric perspective on quantum parameter estimation},
   volume={2},
   ISSN={2639-0213},
   url={http://dx.doi.org/10.1116/1.5119961},
   DOI={10.1116/1.5119961},
   number={1},
   journal={AVS Quantum Science},
   publisher={American Vacuum Society},
   author={Sidhu, Jasminder S. and Kok, Pieter},
   year={2020},
   month=Feb }

@article{zheng2022preparation,
  title={Preparation of metrological states in dipolar-interacting spin systems},
  author={Zheng, Tian-Xing and Li, Anran and Rosen, Jude and Zhou, Sisi and Koppenh{\"o}fer, Martin and Ma, Ziqi and Chong, Frederic T and Clerk, Aashish A and Jiang, Liang and Maurer, Peter C},
  journal={npj Quantum Information},
  volume={8},
  number={1},
  pages={150},
  year={2022},
  publisher={Nature Publishing Group UK London}
}

@article{niu2019universal,
  title={Universal quantum control through deep reinforcement learning},
  author={Niu, Murphy Yuezhen and Boixo, Sergio and Smelyanskiy, Vadim N and Neven, Hartmut},
  journal={npj Quantum Information},
  volume={5},
  number={1},
  pages={33},
  year={2019},
  publisher={Nature Publishing Group UK London}
}

@article{braun2018quantum,
  title={Quantum-enhanced measurements without entanglement},
  author={Braun, Daniel and Adesso, Gerardo and Benatti, Fabio and Floreanini, Roberto and Marzolino, Ugo and Mitchell, Morgan W and Pirandola, Stefano},
  journal={Reviews of Modern Physics},
  volume={90},
  number={3},
  pages={035006},
  year={2018},
  publisher={APS}
}

@article{wang2025tight,
  title={Tight tradeoff relation and optimal measurement for multi-parameter quantum estimation},
  author={Wang, Lingna and Chen, Hongzhen and Yuan, Haidong},
  journal={arXiv preprint arXiv:2504.09490},
  year={2025}
}

@article{yang2019optimal,
  title={Optimal measurements for quantum multiparameter estimation with general states},
  author={Yang, Jing and Pang, Shengshi and Zhou, Yiyu and Jordan, Andrew N},
  journal={Physical Review A},
  volume={100},
  number={3},
  pages={032104},
  year={2019},
  publisher={APS}
}

@article{valeri2020experimental,
  title={Experimental adaptive Bayesian estimation of multiple phases with limited data},
  author={Valeri, Mauro and Polino, Emanuele and Poderini, Davide and Gianani, Ilaria and Corrielli, Giacomo and Crespi, Andrea and Osellame, Roberto and Spagnolo, Nicol{\`o} and Sciarrino, Fabio},
  journal={npj Quantum Information},
  volume={6},
  number={1},
  pages={92},
  year={2020},
  publisher={Nature Publishing Group UK London}
}

@article{demkowicz2017adaptive,
  title={Adaptive quantum metrology under general markovian noise},
  author={Demkowicz-Dobrza{\'n}ski, Rafa{\l} and Czajkowski, Jan and Sekatski, Pavel},
  journal={Physical Review X},
  volume={7},
  number={4},
  pages={041009},
  year={2017},
  publisher={APS}
}

@article{pezze2018quantum,
  title={Quantum metrology with nonclassical states of atomic ensembles},
  author={Pezze, Luca and Smerzi, Augusto and Oberthaler, Markus K and Schmied, Roman and Treutlein, Philipp},
  journal={Reviews of Modern Physics},
  volume={90},
  number={3},
  pages={035005},
  year={2018},
  publisher={APS}
}

@article{le2023variational,
  title={Variational quantum metrology for multiparameter estimation under dephasing noise},
  author={Le, Trung Kien and Nguyen, Hung Q and Ho, Le Bin},
  journal={Scientific Reports},
  volume={13},
  number={1},
  pages={17775},
  year={2023},
  publisher={Nature Publishing Group UK London}
}

@article{yang2020probe,
  title={Probe optimization for quantum metrology via closed-loop learning control},
  author={Yang, Xiaodong and Thompson, Jayne and Wu, Ze and Gu, Mile and Peng, Xinhua and Du, Jiangfeng},
  journal={npj Quantum Information},
  volume={6},
  number={1},
  pages={62},
  year={2020},
  publisher={Nature Publishing Group UK London}
}

@article{43wz-mtrk,
  title = {Quantum metrology in the presence of correlated noise via Markovian embedding},
  author = {Das, Arpan and Demkowicz-Dobrza\ifmmode \acute{n}\else \'{n}\fi{}ski, Rafa\l{}},
  journal = {Phys. Rev. A},
  volume = {113},
  issue = {2},
  pages = {022418},
  numpages = {11},
  year = {2026},
  month = {Feb},
  publisher = {American Physical Society},
  doi = {10.1103/43wz-mtrk},
  url = {https://link.aps.org/doi/10.1103/43wz-mtrk}
}

@article{Yang_2019,
   title={Memory Effects in Quantum Metrology},
   volume={123},
   ISSN={1079-7114},
   url={http://dx.doi.org/10.1103/PhysRevLett.123.110501},
   DOI={10.1103/physrevlett.123.110501},
   number={11},
   journal={Physical Review Letters},
   publisher={American Physical Society (APS)},
   author={Yang, Yuxiang},
   year={2019},
   month=sep }

@article{Sieniawski_2026, doi = {10.1088/1367-2630/ae3edd}, url = {https://doi.org/10.1088/1367-2630/ae3edd}, year = {2026}, month = {feb}, publisher = {IOP Publishing}, volume = {28}, number = {2}, pages = {024502}, author = {Sieniawski, Stanisław and Demkowicz-Dobrzański, Rafał}, title = {Adaptive quantum channel discrimination using methods of quantum metrology}, journal = {New Journal of Physics} }

@article{Bavaresco_2024,
   title={Designing optimal protocols in Bayesian quantum parameter estimation with higher-order operations},
   volume={6},
   ISSN={2643-1564},
   url={http://dx.doi.org/10.1103/PhysRevResearch.6.023305},
   DOI={10.1103/physrevresearch.6.023305},
   number={2},
   journal={Physical Review Research},
   publisher={American Physical Society (APS)},
   author={Bavaresco, Jessica and Lipka-Bartosik, Patryk and Sekatski, Pavel and Mehboudi, Mohammad},
   year={2024},
   month=jun }

@article{Chen_2024,
   title={Simultaneous measurement of multiple incompatible observables and tradeoff in multiparameter quantum estimation},
   volume={10},
   ISSN={2056-6387},
   url={http://dx.doi.org/10.1038/s41534-024-00894-x},
   DOI={10.1038/s41534-024-00894-x},
   number={1},
   journal={npj Quantum Information},
   publisher={Springer Science and Business Media LLC},
   author={Chen, Hongzhen and Wang, Lingna and Yuan, Haidong},
   year={2024},
   month=oct }

@article{Zhang_2026,
   title={Distributed multi-parameter quantum metrology with a superconducting quantum network},
   volume={17},
   ISSN={2041-1723},
   url={http://dx.doi.org/10.1038/s41467-026-68535-9},
   DOI={10.1038/s41467-026-68535-9},
   number={1},
   journal={Nature Communications},
   publisher={Springer Science and Business Media LLC},
   author={Zhang, Jiajian and Wang, Lingna and Hai, Yong-Ju and Zhang, Jiawei and Chu, Ji and Jiang, Ji and Huang, Wenhui and Liang, Yongqi and Qiu, Jiawei and Sun, Xuandong and Tao, Ziyu and Zhang, Libo and Zhou, Yuxuan and Chen, Yuanzhen and Guo, Weijie and Linpeng, Xiayu and Liu, Song and Ren, Wenhui and Zhong, Youpeng and Niu, Jingjing and Yuan, Haidong and Yu, Dapeng},
   year={2026},
   month=jan }

@article{bisio2011minimal,
  title={Minimal computational-space implementation of multiround quantum protocols},
  author={Bisio, Alessandro and D’Ariano, Giacomo Mauro and Perinotti, Paolo and Chiribella, Giulio},
  journal={Physical Review A—Atomic, Molecular, and Optical Physics},
  volume={83},
  number={2},
  pages={022325},
  year={2011},
  publisher={APS}
}

@article{yuan2016sequential,
  title={Sequential feedback scheme outperforms the parallel scheme for Hamiltonian parameter estimation},
  author={Yuan, Haidong},
  journal={Physical review letters},
  volume={117},
  number={16},
  pages={160801},
  year={2016},
  publisher={APS}
}

@article{hayashi2024finding,
  title={Finding the optimal probe state for multiparameter quantum metrology using conic programming},
  author={Hayashi, Masahito and Ouyang, Yingkai},
  journal={npj Quantum Information},
  volume={10},
  number={1},
  pages={111},
  year={2024},
  publisher={Nature Publishing Group UK London}
}

@article{chiribella2009theoretical,
  title={Theoretical framework for quantum networks},
  author={Chiribella, Giulio and D’Ariano, Giacomo Mauro and Perinotti, Paolo},
  journal={Physical Review A—Atomic, Molecular, and Optical Physics},
  volume={80},
  number={2},
  pages={022339},
  year={2009},
  publisher={APS}
}

@article{PhysRevLett.127.110402,
  title = {Quantum Theory Cannot Violate a Causal Inequality},
  author = {Purves, Tom and Short, Anthony J.},
  journal = {Phys. Rev. Lett.},
  volume = {127},
  issue = {11},
  pages = {110402},
  numpages = {6},
  year = {2021},
  month = {Sep},
  publisher = {American Physical Society},
  doi = {10.1103/PhysRevLett.127.110402},
  url = {https://link.aps.org/doi/10.1103/PhysRevLett.127.110402}
}

@article{chiribella2008quantum,
  title={Quantum circuit architecture},
  author={Chiribella, Giulio and D’Ariano, G Mauro and Perinotti, Paolo},
  journal={Physical review letters},
  volume={101},
  number={6},
  pages={060401},
  year={2008},
  publisher={APS}
}

@misc{pang2026,
     title={A Unified Conic-Programming Framework for Multiparameter Quantum Metrology with Indefinite Causal Order}, 
     author={Wenjie Wei and Shengshi Pang},
     year={2026},
}

@misc{zhou2026unifiedcomputableapproachoptimal,
      title={Unified and computable approach to optimal strategies for multiparameter estimation}, 
      author={Zhao-Yi Zhou and Da-Jian Zhang},
      year={2026},
      eprint={2603.06244},
      archivePrefix={arXiv},
      primaryClass={quant-ph},
      url={https://arxiv.org/abs/2603.06244}, 
}

@article{kurdzialek2023using,
  title={Using adaptiveness and causal superpositions against noise in quantum metrology},
  author={Kurdzia{\l}ek, Stanis{\l}aw and G{\'o}recki, Wojciech and Albarelli, Francesco and Demkowicz-Dobrza{\'n}ski, Rafa{\l}},
  journal={Physical Review Letters},
  volume={131},
  number={9},
  pages={090801},
  year={2023},
  publisher={APS}
}

@article{araujo2014computational,
  title={Computational advantage from quantum-controlled ordering of gates},
  author={Ara{\'u}jo, Mateus and Costa, Fabio and Brukner, {\v{C}}aslav},
  journal={Physical review letters},
  volume={113},
  number={25},
  pages={250402},
  year={2014},
  publisher={APS}
}

@article{wechs2021quantum,
  title={Quantum circuits with classical versus quantum control of causal order},
  author={Wechs, Julian and Dourdent, Hippolyte and Abbott, Alastair A and Branciard, Cyril},
  journal={PRX Quantum},
  volume={2},
  number={3},
  pages={030335},
  year={2021},
  publisher={APS}
}

@article{rubino2017experimental,
  title={Experimental verification of an indefinite causal order},
  author={Rubino, Giulia and Rozema, Lee A and Feix, Adrien and Ara{\'u}jo, Mateus and Zeuner, Jonas M and Procopio, Lorenzo M and Brukner, {\v{C}}aslav and Walther, Philip},
  journal={Science advances},
  volume={3},
  number={3},
  pages={e1602589},
  year={2017},
  publisher={American Association for the Advancement of Science}
}

@article{chiribella2013quantum,
  title={Quantum computations without definite causal structure},
  author={Chiribella, Giulio and D’Ariano, Giacomo Mauro and Perinotti, Paolo and Valiron, Benoit},
  journal={Physical Review A—Atomic, Molecular, and Optical Physics},
  volume={88},
  number={2},
  pages={022318},
  year={2013},
  publisher={APS}
}

@misc{hu2024controlincompatibilitymultiparameterquantum,
      title={Control incompatibility in multiparameter quantum metrology}, 
      author={Zhiyao Hu and Shilin Wang and Linmu Qiao and Takuya Isogawa and Changhao Li and Yu Yang and Guoqing Wang and Haidong Yuan and Paola Cappellaro},
      year={2024},
      eprint={2411.18896},
      archivePrefix={arXiv},
      primaryClass={quant-ph},
      url={https://arxiv.org/abs/2411.18896}, 
}

@article{guhne2023colloquium,
  title={Colloquium: Incompatible measurements in quantum information science},
  author={G{\"u}hne, Otfried and Haapasalo, Erkka and Kraft, Tristan and Pellonp{\"a}{\"a}, Juha-Pekka and Uola, Roope},
  journal={Reviews of Modern Physics},
  volume={95},
  number={1},
  pages={011003},
  year={2023},
  publisher={APS}
}

@article{matsuzaki2011magnetic,
  title={Magnetic field sensing beyond the standard quantum limit under the effect of decoherence},
  author={Matsuzaki, Yuichiro and Benjamin, Simon C and Fitzsimons, Joseph},
  journal={Physical Review A—Atomic, Molecular, and Optical Physics},
  volume={84},
  number={1},
  pages={012103},
  year={2011},
  publisher={APS}
}

@article{moreau2019imaging,
  title={Imaging with quantum states of light},
  author={Moreau, Paul-Antoine and Toninelli, Ermes and Gregory, Thomas and Padgett, Miles J},
  journal={Nature Reviews Physics},
  volume={1},
  number={6},
  pages={367--380},
  year={2019},
  publisher={Nature Publishing Group UK London}
}

@article{hou2020minimal,
  title={Minimal tradeoff and ultimate precision limit of multiparameter quantum magnetometry under the parallel scheme},
  author={Hou, Zhibo and Zhang, Zhao and Xiang, Guo-Yong and Li, Chuan-Feng and Guo, Guang-Can and Chen, Hongzhen and Liu, Liqiang and Yuan, Haidong},
  journal={Physical Review Letters},
  volume={125},
  number={2},
  pages={020501},
  year={2020},
  publisher={APS}
}

@article{chao2016fisher,
  title={Fisher information theory for parameter estimation in single molecule microscopy: tutorial},
  author={Chao, Jerry and Sally Ward, E and Ober, Raimund J},
  journal={Journal of the Optical Society of America A},
  volume={33},
  number={7},
  pages={B36--B57},
  year={2016},
  publisher={Optical Society of America}
}

@article{dorfman2016nonlinear,
  title={Nonlinear optical signals and spectroscopy with quantum light},
  author={Dorfman, Konstantin E and Schlawin, Frank and Mukamel, Shaul},
  journal={Reviews of Modern Physics},
  volume={88},
  number={4},
  pages={045008},
  year={2016},
  publisher={APS}
}

@article{hou2021super,
  title={“Super-Heisenberg” and Heisenberg scalings achieved simultaneously in the estimation of a rotating field},
  author={Hou, Zhibo and Jin, Yan and Chen, Hongzhen and Tang, Jun-Feng and Huang, Chang-Jiang and Yuan, Haidong and Xiang, Guo-Yong and Li, Chuan-Feng and Guo, Guang-Can},
  journal={Physical Review Letters},
  volume={126},
  number={7},
  pages={070503},
  year={2021},
  publisher={APS}
}

@article{Albarelli_2019,
   title={Evaluating the Holevo Cramér-Rao Bound for Multiparameter Quantum Metrology},
   volume={123},
   ISSN={1079-7114},
   url={http://dx.doi.org/10.1103/PhysRevLett.123.200503},
   DOI={10.1103/physrevlett.123.200503},
   number={20},
   journal={Physical Review Letters},
   publisher={American Physical Society (APS)},
   author={Albarelli, Francesco and Friel, Jamie F. and Datta, Animesh},
   year={2019},
   month=nov }

@article{dulian2025qmetro++,
  title={QMetro++--Python optimization package for large scale quantum metrology with customized strategy structures},
  author={Dulian, Piotr and Kurdzia{\l}ek, Stanis{\l}aw and Demkowicz-Dobrza{\'n}ski, Rafa{\l}},
  journal={arXiv preprint arXiv:2506.16524},
  year={2025}
}

@article{albarelli2022probe,
  title={Probe incompatibility in multiparameter noisy quantum metrology},
  author={Albarelli, Francesco and Demkowicz-Dobrza{\'n}ski, Rafa{\l}},
  journal={Physical Review X},
  volume={12},
  number={1},
  pages={011039},
  year={2022},
  publisher={APS}
}

@article{Chen_2026,
   title={Optimal Quantum Metrology under Energy Constraints},
   volume={136},
   ISSN={1079-7114},
   url={http://dx.doi.org/10.1103/6ghs-frtx},
   DOI={10.1103/6ghs-frtx},
   number={7},
   journal={Physical Review Letters},
   publisher={American Physical Society (APS)},
   author={Chen, Longyun and Yang, Yuxiang},
   year={2026},
   month=feb 
}

@article{Yamagata_2013,
   title={Quantum local asymptotic normality based on a new quantum likelihood ratio},
   volume={41},
   ISSN={0090-5364},
   url={http://dx.doi.org/10.1214/13-AOS1147},
   DOI={10.1214/13-aos1147},
   number={4},
   journal={The Annals of Statistics},
   publisher={Institute of Mathematical Statistics},
   author={Yamagata, Koichi and Fujiwara, Akio and Gill, Richard D.},
   year={2013},
   month=Aug }

\appendix
\onecolumngrid
\section{Expressions for Quantum Strategies}
\label{appendix:Expressions for Quantum Strategies}
Here we introduce the general quantum strategies including definite causal order strategies including sequential and parallel and indefinite causal order ones (ICO) like quantum SWITCH and causal superposition, as well as their mathematical formulism.

\subsection{Definite causal order Strategies}

Given parameterized quantum channels with total Choi-Jamiołkowski (CJ) operator $N_{\bm{\theta}}=\bigotimes_{i=1}^{N} E^{(i)}_{\bm{\theta}}\in \mathcal{L}(\mathcal{H}_{S})$, a sequential strategy is represented by a positive semidefinite operator $P$ satisfying the recursive quantum-comb normalization conditions
\begin{equation}
\begin{aligned}
&P=P^{(N)},\
&\mathrm{Tr}_{2k-1}[P^{(k)}]=\mathcal{I}_{2k-2}\otimes P^{(k-1)},~k=N,\dots,2,\
&\mathrm{Tr}[P^{(1)}]=1,
\end{aligned}
\end{equation}
where $P^{(N)}\in \mathcal{L}(\mathcal{H}_{\bar{S}})=\mathcal{L}(\bigotimes_{i=1}^{2N-1}\mathcal{H}_{i})$ is an $N$-step quantum comb. These constraints encode the causal ordering of the $N$ channel uses and ensure that each intermediate operation is trace preserving after tracing over the corresponding input system.

A parallel strategy corresponds to a single-step quantum comb. It is specified by an operator $P$ satisfying
\begin{equation}
\begin{aligned}
&P=P^{(1)},\text{ Tr}[P^{(1)}]=1,
\end{aligned}
\end{equation}
where $P^{(1)}$ is defined on $\mathcal{L}(\mathcal{H}_{\bar{S}})=\mathcal{L}(\bigotimes_{i=1}^{N}\mathcal{H}_{2i-1})$. In this case, the initial probe state is optimized, possibly with the assistance of ancillary systems, while no intermediate controls are inserted during the evolution. Equivalently, the $N$ channel uses are treated as a single composite channel from $\mathcal{L}(\bigotimes_{i=1}^{N}\mathcal{H}_{2i-1})$ to $\mathcal{L}(\bigotimes_{i=1}^{N}\mathcal{H}_{2i})$. In this sense, parallel strategies form a restricted subclass of sequential strategies.

Both strategy classes are defined with a trivial input space $\mathcal{H}_0=\mathbb{C}$. This convention incorporates state preparation into the strategy itself, so that the strategy does not require an external input state. In particular, the first sub-comb $P^{(1)}$ is precisely the reduced probe state. For sequential strategies, the intermediate controls can be recovered from the higher-order sub-combs $P^{(k)}$, $k=N,\dots,2$, following the realization procedure introduced in Theorems 1 and 2 of Ref.~\cite{bisio2011minimal}.

\subsection{ICO Strategies}
An $N$-partite quantum SWITCH strategy coherently combines the $N!$ possible causal orders of the channels. In the present representation, it is described by a set of positive semidefinite constraints
\begin{equation}\begin{aligned}
    &P_{\text{ICO}}=\sum_\pi q_\pi P_\pi\otimes\mathcal{I}_{2\pi(N)},\\
    &P_\pi= \rho^\pi_{2\pi(1)-1}(\bigotimes_{i=1}^{N-1}|I\rrangle \llangle I|_{2\pi(i),2\pi(i+1)-1} )\otimes \mathcal{I}_{2\pi(N)},\\
    &\sum_{\pi\in S_N}q_\pi=1,q_\pi\ge0,\rho_{2\pi(1)-1}\ge 0,~\mathrm{Tr}[\rho_{2\pi(1)-1}]=1,~\pi\in S_N,
\end{aligned}\end{equation}
where each permutation $\pi$ belongs to the symmetric group $S_N$ of degree $N$. Within each causal order, the channels are connected by identity maps and no intermediate controls are applied. Therefore, optimizing the quantum SWITCH strategy is equivalent to optimizing the probe state in each order sector, together with the weights associated with different causal orders.

In order to consider intermediate controls, the causal order superposition, which is the linear combination of sequential strategies is proposed. The constraints of a causal superposition strategy is given as
\begin{equation}
    \begin{aligned}
            &P_{\text{ICO}}=\sum_\pi q_\pi P_\pi\otimes\mathcal{I}_{2\pi(N)},~P^\pi=P^{\pi,(N)}\otimes \mathcal{I}_{2\pi (N)},~\sum_{\pi}q_\pi=1,\\
            &\text{Tr}_{2\pi(k)-1}[P^{\pi,(k)}]=\mathcal{I}_{2\pi(k)-2}\otimes P^{\pi,(k-1)},~k=N,\dots,2,\\
                &\text{Tr}[P^{\pi,(1)}]=1,~q_\pi\ge0,~\pi\in S_N,
    \end{aligned} 
\end{equation}
where each component strategy follows the order
$\mathcal{E}^{\pi(1)}_{\bm{\theta}} \to \mathcal{E}^{\pi(2)}_{\bm{\theta}}\to\cdots \to \mathcal{E}^{\pi(N)}_{\bm{\theta}}$. Here $\mathcal{E}^{(k)}_{\bm{\theta}}$ denotes the channel from $\mathcal{L}(\mathcal{H}_{2k-1})$ to $\mathcal{L}(\mathcal{H}_{2k})$. Compared with the quantum SWITCH form above, causal superposition strategies allow nontrivial intermediate controls within each order sector, and hence provide a broader class of indefinite-causal-order protocols.

A general indefinite-causal-order strategy is defined as $\tilde{P}_{\text{ICO}}\in \mathcal{H}_S\otimes \mathcal{H}_F$, where $\mathcal{H}_F$ denotes the global output space for the strategy
\begin{equation}
\label{plain_condition_ICO}
\begin{aligned}
    &\mathsf{rank}(\tilde P_{\text{ICO}})=1,~ \tilde P_{\text{ICO}}\ge0,~\tilde P_{\text{ICO}}\star(\bigotimes_{i=1}^N E^{(i)}_{\bm{\theta}})\ge0,~\text{Tr}[\tilde P_{\text{ICO}}\star(\bigotimes_{i=1}^N E^{(i)}_{\bm{\theta}})]=1,\\
\end{aligned}
\end{equation}
for $N=2$, the condition can be formulated as a more programmable-friendly expression with $P_{\text{ICO}}=\text{Tr}_F[\tilde P_{\text{ICO}}]\in\mathcal{H}_S$:
\begin{equation}
\label{N=2_condition_ICO}
\begin{aligned}
    &P_{\text{ICO}}\ge 0, ~\text{Tr}[P_{\text{ICO}}]=d_2*d_4,\\
    &\prescript{}{1,2}{P_{\text{ICO}}}=\prescript{}{1,2,4}{P_{\text{ICO}}},\\
    &\prescript{}{3,4}{P_{\text{ICO}}}=\prescript{}{2,3,4}{P_{\text{ICO}}},\\
    &P_{\text{ICO}}=\prescript{}{2}{P_{\text{ICO}}}+\prescript{}{4}{P_{\text{ICO}}}-\prescript{}{2,4}{P_{\text{ICO}}},\\
\end{aligned}
\end{equation}
where $\prescript{}{x}{P}=\frac{\mathcal{I}_x}{d_x}\otimes \text{Tr}_{x}[P]$, which can be proved equivalent to \ref{plain_condition_ICO} (see \cite[Appendix~B]{araujo2015witnessing}).

\section{Equivalence between WMSE and $Q^{P}$}
\label{appendix:Proof for Q}

We first show how the weighted mean-square error can be rewritten in the
operator form used in the main text. Define
\(\Delta\bm{\theta}(x):=\hat{\bm{\theta}}(x)-\bm{\theta}\).
By Born's rule
\(p(x|\bm{\theta})=\mathrm{Tr}(\rho_{\bm{\theta}}M_x)\), with
\(\sum_x M_x=\mathcal{I}\), the weighted mean-square error is
\begin{align}
    \mathrm{Tr}[WV]
    &=\sum_x p(x|\bm\theta)\,
    \mathrm{Tr}\!\left[
        W\Delta\bm{\theta}(x)\Delta\bm{\theta}(x)^T
    \right] \nonumber\\
    &=\sum_x p(x|\bm\theta)\,
    \Delta\bm{\theta}(x)^T W\Delta\bm{\theta}(x).
\end{align}
For a locally unbiased estimator,
\(\mathbb{E}_{\bm{\theta}}[\Delta\bm{\theta}]=0\).

To rewrite the weighted mean-square error, introduce a
\((d+1)\)-dimensional reference space \(\mathcal R^{d+1}\) with basis
\(\{|0\rangle,|1\rangle,\dots,|d\rangle\}\), and define
\begin{equation}
    X :=
    \sum_x
    \begin{bmatrix}
        1 & \Delta\bm{\theta}(x)^T \\
        \Delta\bm{\theta}(x) &
        \Delta\bm{\theta}(x)\Delta\bm{\theta}(x)^T
    \end{bmatrix}
    \otimes M_x
    \in \mathcal{L}(H_{\mathcal R}\otimes\mathcal{H}_O).
\end{equation}
If
\(\mathbf v_x:=[1;\Delta\bm{\theta}(x)]\), then
\(X=\sum_x\mathbf v_x\mathbf v_x^T\otimes M_x\).
We further lift \(W\in\mathcal{L}(\mathbb{R}^d)\) to
\(\mathcal{L}(\mathbb{R}^{d+1})\) as
\begin{equation}
    \bar W :=
    0\oplus W.
\end{equation}
Therefore,
\begin{align}
    \mathrm{Tr}\!\left[
        (\bar W\otimes\rho_{\bm{\theta}})X
    \right]
    &=
    \sum_x
    \mathrm{Tr}[\rho_{\bm{\theta}}M_x]\,
    \mathrm{Tr}[\bar W\mathbf v_x\mathbf v_x^T]
    \nonumber\\
    &=
    \sum_x p(x|\bm{\theta})\,
    \Delta\bm{\theta}(x)^T
    W\Delta\bm{\theta}(x)
    \nonumber\\
    &=\mathrm{Tr}[WV].
\end{align}
Consequently,
\begin{equation}
    \mathrm{Tr}[WV]
    =
    \mathrm{Tr}\!\left[
        (\bar W\otimes\rho_{\bm{\theta}})X
    \right]
    =
    \mathrm{Tr}\!\left[
        \bigl(\bar W\otimes(N_{\bm{\theta}}*\tilde P)\bigr)X
    \right].
\end{equation}
This yields
\begin{equation}
    Q^{P}
    :=
    \min_{P\in\mathcal P,\;X}
    \mathrm{Tr}\!\left[
        \bigl(\bar W\otimes(N_{\bm{\theta}}*\tilde P)\bigr)X
    \right].
\end{equation}

The constraints take a linear form in \(X\). The normalization of the POVM gives
\begin{equation}
    \mathrm{Tr}_{\mathcal R}\!\left[(|0\rangle\langle0|_{\mathcal R}\otimes \mathcal{I}_O)X\right]=\mathcal{I}_O.
\end{equation}
Moreover, since $\frac12\mathrm{Tr}\!\left[(|0\rangle\langle i|+|i\rangle\langle0|)\mathbf v_x\mathbf v_x^T\right]=\hat\theta_i(x)$, the first-order l.u. condition is equivalently written as
\begin{equation}
    \frac12\mathrm{Tr}\!\left[\left((|0\rangle\langle i|_{\mathcal R}+|i\rangle\langle0|_{\mathcal R})\otimes(\partial_{\theta_j}N_{\bm\theta}*\tilde P)\right)X\right]
    =\delta_{ij}.
\end{equation}
These are precisely the constraints used in the main text.

\section{Expressions for Cones $\mathcal{C}_k$}
\label{appendix:Expressions for Cones}

We collect here the explicit expressions for the cones $\mathcal{C}_k$, $k=1,2,3,4$, associated with the tight, Nagaoka--Hayashi, SLD, and Holevo bounds, following Ref.~\cite{hayashi2023tight}. These cones specify different feasible sets for the auxiliary operator used to encode the measurement and estimator, and therefore determine the corresponding multi-parameter precision bounds.

For the tight bound,
\begin{equation}
    \mathcal{C}_1
    :=\mathrm{conv}\left\{W\otimes Z \bigg| W\in \mathcal{L}(\mathcal{R})_{rs,+}, Z\in \mathcal{B}(\mathcal{H}_O)_+\right\},
\end{equation}
namely the convex hull of tensor products between real symmetric positive semidefinite matrices on $\mathcal{R}=\mathbb{R}^{d+1}$ and bounded positive semidefinite operators on $\mathcal{H}_O$. 

For the Nagaoka-Hayashi bound, the cone $\mathcal{C}_2 = (\mathbb{C}^{d+1} \otimes \mathcal{H})_P \cap \mathcal{B}$ is the intersection of all the positive semideifinite operators on $\mathbb{C}^{d+1} \otimes \mathcal{H}$ with the set $\mathcal{B}$ defined as:
\begin{equation}
\begin{aligned}
    \mathcal{B}:=\left\{\sum_{j,k=0}^d \ket{k}\bra{j}\otimes X_{k,j} \bigg|X_{k,j} \text{ bounded, } X_{k,j}=X_{k,j}^\dagger, X_{k,j}=X_{j,k}\right\},
\end{aligned}
\end{equation}
where $X_{k,j}$ are bounded Hermitians on $\mathcal{H}$. 

The SLD bound corresponds to the cone $\mathcal{C}_3 = (\mathbb{C}^{d+1} \otimes \mathcal{H})_P \cap \mathcal{B}^{\prime\prime}$ where:
\begin{equation}
\begin{aligned}   \mathcal{B}^{\prime\prime}:=\left\{\sum_{j,k=0}^d \ket{k}\bra{j}\otimes X_{k,j} \bigg|X_{k,0} \text{ bounded, } X_{k,0}=X_{k,0}^\dagger, X_{k,j}=X_{j,k}^\dagger\right\}.
\end{aligned}
\end{equation}
For the Holevo bound $\mathcal{C}_4$, the set $\mathcal{B}_\Omega^{\prime\prime}$ is further restricted by a linear constraint with respect to $\Omega$:
\begin{equation}
    \mathrm{Tr}\left[\Omega((\ket{i}\bra{j}_{\mathcal{R}}-\ket{j}\bra{i}_{\mathcal{R}})\otimes T_{\bm{\theta}})\right]=0,
\end{equation}
i.e. $\mathcal{C}_4:=(\mathbb{C}^{d+1} \otimes \mathcal{H})_P \cap\mathcal{B}_\Omega^{\prime\prime}$. 

These cones obey the inclusion relation
\begin{equation}
\mathcal{C}_1 \subseteq \mathcal{C}_2 \subseteq \mathcal{C}_4 \subseteq \mathcal{C}_3,
\end{equation}
which corresponds to the hierarchy of multi-parameter bounds,
\begin{equation}
C^{\text{tight}}\ge C^{\text{NH}}\ge C^{\text{H}}\ge C^{\text{SLD}}.
\end{equation}
In multi-parameter estimation, the Holevo bound is tight for pure states, so that $C^{\text{tight}}= C^{\text{NH}}= C^{\text{H}}$, whereas gaps may appear for mixed states.

All the above cones are convex. However, the separable cone $\mathcal{C}_1$ is not, in general, semidefinite representable. Therefore, in the present work we focus on SDP formulations of the Holevo, Nagaoka--Hayashi, and SLD bounds, while the upper and lower bounds of $C^{\text{tight}}$ can be evaluated using SDP relaxations \cite{hayashi2023tight,hayashi2024finding}.

\section{Proof of $S_k^P= Q_k^P$}
\label{appendix:proof of theorem 1}

We now prove the equivalence between the original formulation $Q_k^P$ and the conic formulation $S_k^P$ for $k=2,3,4$. The proof consists of two directions. First, we show that every feasible solution of $Q_k^P$ induces a feasible solution of $S_k^P$ with the same objective value. Second, we show the converse by constructing, from any feasible solution of $S_k^P$, a feasible measurement-estimator operator and a purified strategy for the original formulation.

\subsection{Proof of $S_k^P\le Q_k^P$}
\label{appendix:Proof of S_k^P<=Q_k^P}
Suppose first that $\tilde P^*$ is a feasible purified strategy for $Q_k^P$, and that
$X^*\in\mathcal C_k$ is the corresponding feasible measurement-estimator operator. Define
\begin{equation}
    \Omega
    =
    \mathrm{Tr}_{O_M}
    \left[
        \left(
            \tilde P^*
            \otimes
            \mathcal I_{\mathcal R,O_S}
        \right)
        \left(
            \mathcal I_{\overline S}
            \otimes
            X^*
        \right)
    \right].
    \label{eq:Omega_forward_general}
\end{equation}
Equivalently,
\begin{equation}
    \Omega
    =
    \left(
        \mathcal I_{\mathcal R}
        \otimes
        \Phi_{\tilde P^*}
    \right)(X^*),
\end{equation}
where the strategy contraction
$\Phi_{\tilde P^*}$ maps operators on
$\mathcal H_O=\mathcal H_{O_S}\otimes\mathcal H_{O_M}$
to operators on
$\mathcal H_S=\mathcal H_{O_S}\otimes\mathcal H_{\overline S}$.
Under the Choi-link convention used in this work, $\Phi_{\tilde P^*}$ is completely positive. Therefore
$\mathcal I_{\mathcal R}\otimes\Phi_{\tilde P^*}$ preserves positive semidefiniteness.

We first verify the conic constraint. For the tight-bound cone, if
\begin{equation}
    X^*
    \in
    \mathcal C_1
    =
    \mathrm{conv}
    \left\{
        W_\mu\otimes Z_\mu
        \,\middle|\,
        W_\mu\in\mathcal L(\mathcal R)_{rs,+},
        \;
        Z_\mu\in\mathcal B(\mathcal H_O)_+
    \right\},
\end{equation}
then $X^*$ admits a decomposition
\begin{equation}
    X^*
    =
    \sum_\mu
    W_\mu
    \otimes
    Z_\mu,
    \qquad
    W_\mu\in\mathcal L(\mathcal R)_{rs,+},
    \quad
    Z_\mu\in\mathcal B(\mathcal H_O)_+ .
\end{equation}
Applying the contraction gives
\begin{equation}
    \Omega
    =
    \sum_\mu
    W_\mu
    \otimes
    \Phi_{\tilde P^*}(Z_\mu).
\end{equation}
Since $\Phi_{\tilde P^*}$ is positive,
$\Phi_{\tilde P^*}(Z_\mu)\in\mathcal B(\mathcal H_S)_+$ for every
$Z_\mu\ge0$. This does not require $\Phi_{\tilde P^*}$ to be surjective onto
$\mathcal B(\mathcal H_S)_+$. It is sufficient that each term in the separable decomposition is mapped to another product of positive operators. Hence
\begin{equation}
    \Omega
    \in
    \mathrm{conv}
    \left\{
        W\otimes Z
        \,\middle|\,
        W\in\mathcal L(\mathcal R)_{rs,+},
        \;
        Z\in\mathcal B(\mathcal H_S)_+
    \right\}
    =
    \tilde{\mathcal{C}}_1.
\end{equation}

For the Nagaoka--Hayashi and SLD cones, write
\begin{equation}
    X^*
    =
    \sum_{j,k=0}^{d}
    \ket{k}\bra{j}
    \otimes
    X^*_{k,j}.
\end{equation}
Then
\begin{equation}
    \Omega
    =
    \sum_{j,k=0}^{d}
    \ket{k}\bra{j}
    \otimes
    \Phi_{\tilde P^*}(X^*_{k,j}).
\end{equation}
Since $\mathcal I_{\mathcal R}\otimes\Phi_{\tilde P^*}$ is positive, it preserves the global positive-semidefinite constraint on
$\mathbb C^{d+1}\otimes\mathcal H_O$. Moreover, $\Phi_{\tilde P^*}$ is Hermiticity preserving. Hence the block constraints defining $\mathcal B$ and $\mathcal B^{\prime\prime}$ are preserved. In particular, if
$X^*_{k,j}=X^{*\dagger}_{k,j}$ and
$X^*_{k,j}=X^*_{j,k}$, then
\begin{equation}
    \Phi_{\tilde P^*}(X^*_{k,j})
    =
    \Phi_{\tilde P^*}(X^*_{k,j})^\dagger,
    \qquad
    \Phi_{\tilde P^*}(X^*_{k,j})
    =
    \Phi_{\tilde P^*}(X^*_{j,k}),
\end{equation}
so the Nagaoka--Hayashi cone constraint is preserved. Similarly, if
$X^*_{k,0}=X^{*\dagger}_{k,0}$ and
$X^*_{k,j}=X^{*\dagger}_{j,k}$, then
\begin{equation}
    \Phi_{\tilde P^*}(X^*_{k,0})
    =
    \Phi_{\tilde P^*}(X^*_{k,0})^\dagger,
    \qquad
    \Phi_{\tilde P^*}(X^*_{k,j})
    =
    \Phi_{\tilde P^*}(X^*_{j,k})^\dagger,
\end{equation}
so the SLD cone constraint is preserved. Therefore,
\begin{equation}
    X^*\in\mathcal C_k
    \quad\Longrightarrow\quad
    \Omega\in \tilde{\mathcal {C}}_k,
    \qquad
    k=2,3.
\end{equation}

For the Holevo cone, the above argument preserves the positive-semidefinite constraint and the $\mathcal B^{\prime\prime}$ block constraint. It remains to verify the additional Holevo linear constraint. Let
\begin{equation}
    \Lambda_{ij}
    :=
    \ket{i}\bra{j}_{\mathcal R}
    -
    \ket{j}\bra{i}_{\mathcal R}.
\end{equation}
The output state associated with $\tilde P^*$ is
\begin{equation}
    \rho^*_{\bm\theta}
    =
    N_{\bm\theta}*\tilde P^*
    =
    \mathrm{Tr}_{\overline S}
    \left[
        \left(
            \tilde P^*
            \otimes
            \mathcal I_S
        \right)
        \left(
            \mathcal I_{O_M}
            \otimes
            T_{\bm\theta}
        \right)
    \right].
\end{equation}
Using the definition of $\Omega$ and the link-product identity, we obtain
\begin{equation}
    \mathrm{Tr}
    \left[
        \Omega
        \left(
            \Lambda_{ij}
            \otimes
            T_{\bm\theta}
        \right)
    \right]
    =
    \mathrm{Tr}
    \left[
        X^*
        \left(
            \Lambda_{ij}
            \otimes
            \rho^*_{\bm\theta}
        \right)
    \right].
\end{equation}
Thus, if $X^*$ satisfies the Holevo constraint
\begin{equation}
    \mathrm{Tr}
    \left[
        X^*
        \left(
            \Lambda_{ij}
            \otimes
            \rho^*_{\bm\theta}
        \right)
    \right]
    =
    0,
\end{equation}
then $\Omega$ satisfies
\begin{equation}
    \mathrm{Tr}
    \left[
        \Omega
        \left(
            \Lambda_{ij}
            \otimes
            T_{\bm\theta}
        \right)
    \right]
    =
    0.
\end{equation}
Therefore $X^*\in\mathcal C_4$ implies
$\Omega\in\tilde{\mathcal {C}}_4$.

The normalization and local-unbiasedness constraints are transformed by the same contraction. The POVM normalization condition gives
\begin{equation}
    \mathrm{Tr}_{\mathcal R}
    \left[
        \Omega
        \left(
            \ket{0}\bra{0}_{\mathcal R}
            \otimes
            \mathcal I_S
        \right)
    \right]
    =
    P^*\otimes I_{O_S},
\end{equation}
where $P^*=\mathrm{Tr}_{O_M}[\tilde P^*]$ is an admissible strategy operator. The local-unbiasedness constraints become
\begin{equation}
    \frac{1}{2}
    \mathrm{Tr}
    \left[
        \Omega
        \left(
            \ket{0}\bra{i}_{\mathcal R}
            +
            \ket{i}\bra{0}_{\mathcal R}
        \right)
        \otimes
        F_j
    \right]
    =
    \delta_{ij},
\end{equation}
with $F_j=\partial T_{\bm\theta}/\partial\theta^j$. Therefore $\Omega$ is feasible for $S_k^P$.

The objective values also coincide. Indeed,
\begin{equation}
\begin{aligned}
    \mathrm{Tr}
    \left[
        \Omega
        (W\otimes T_{\bm\theta})
    \right]
    &=
    \mathrm{Tr}
    \left[
        \left(
            W
            \otimes
            \mathrm{Tr}_{\overline S}
            \left[
                \left(
                    \tilde P^*
                    \otimes
                    \mathcal I_S
                \right)
                \left(
                    \mathcal I_{O_M}
                    \otimes
                    T_{\bm\theta}
                \right)
            \right]
        \right)
        X^*
    \right]\\
    &=
    \mathrm{Tr}
    \left[
        (W\otimes\rho^*_{\bm\theta})
        X^*
    \right].
\end{aligned}
\end{equation}
Hence every feasible point of $Q_k^P$ induces a feasible point of $S_k^P$ with the same objective value, and therefore
\begin{equation}
    S_k^P\le Q_k^P.
\end{equation}

\subsection{Proof of $S_k^P\ge Q_k^P$}
\label{appendix:Proof of S_k^P>=Q_k^P}

We now prove the converse direction. Let $\Omega^*$ be a feasible solution of $S_k^P$, and let $P^*$ be the strategy operator determined by condition $\mathrm{(i)}$,
\begin{equation}
    \mathrm{Tr}_{\mathcal R}
    \left[
        \Omega^*
        \left(
            \ket{0}\bra{0}_{\mathcal R}
            \otimes
            \mathcal I_S
        \right)
    \right]
    =
    P^*\otimes I_{O_S}.
\end{equation}
If $P^*$ is not strictly positive on its support, we use a standard regularization. Let
\begin{equation}
    D_\epsilon
    =
    \sqrt{1-\epsilon}\ket{0}\bra{0}
    +
    \frac{1}{\sqrt{1-\epsilon}}
    \sum_{i=1}^{d}
    \ket{i}\bra{i},
    \qquad
    0<\epsilon<1.
\end{equation}·ne
\begin{equation}
    \Omega_\epsilon
    =
    (D_\epsilon\otimes\mathcal I_S)
    \Omega^*
    (D_\epsilon\otimes\mathcal I_S)
    +
    \frac{\epsilon}{\prod^{2N}_{k=1}d_k}
    \ket{0}\bra{0}_{\mathcal R}
    \otimes
    I_{S}.
\end{equation}
The local-unbiasedness constraints are unchanged because
$D_\epsilon$ rescales the $|0\rangle\langle i|$ and
$|i\rangle\langle0|$ blocks by reciprocal factors, and the added term has support only on the $|0\rangle\langle0|$ block. The conic constraint is preserved for all four cones: for $\mathcal C_1$ the added term is separable, for $\mathcal C_2$ and $\mathcal C_3$ the block constraints are preserved, and for $\mathcal C_4$ the added term does not contribute to the antisymmetric Holevo constraint. Moreover, the strategy operator associated with $\Omega_\epsilon$ is
\begin{equation}
    P_\epsilon
    =
    (1-\epsilon)P^*
    +
    \frac{\epsilon}{\prod_{k=1}^Nd_k}I_{\bar{S}},
\end{equation}
which is strictly positive. Since $\Omega_\epsilon\to\Omega^*$ as $\epsilon\to0^+$, it suffices, by continuity of the objective function, to construct a feasible point of $Q_k^P$ from $\Omega_\epsilon$.

For fixed $\epsilon>0$, take the spectral decomposition
\begin{equation}
    P_\epsilon
    =
    \sum_j
    p_j
    \ket{P_j}\bra{P_j},
    \qquad
    p_j>0.
\end{equation}
Choose an ancillary output space $\mathcal H_{O_M}$ with dimension at least
$\dim(\mathcal H_{\overline S})$, and choose a unitary isomorphism
$U:\mathcal H_{\overline S}\to\mathcal H_{O_M}$ such that
\begin{equation}
    \ket{P_j^\prime}
    =
    U\ket{P_j}
\end{equation}
forms an orthonormal set in $\mathcal H_{O_M}$. Define a purification of $P_\epsilon$ by
\begin{equation}
    \tilde P_\epsilon
    =
    \ket{\tilde P_\epsilon}\bra{\tilde P_\epsilon},
    \qquad
    \ket{\tilde P_\epsilon}
    =
    \sum_j
    p_j^{1/2}
    \ket{P_j}
    \ket{P_j^\prime}.
\end{equation}
Then define the lifted operator
\begin{equation}
    X_\epsilon
    =
    \left[
        \mathcal I_{\mathcal R,O_S}
        \otimes
        U P_\epsilon^{-1/2}
    \right]
    \Omega_\epsilon
    \left[
        \mathcal I_{\mathcal R,O_S}
        \otimes
        P_\epsilon^{-1/2} U^\dagger
    \right].
    \label{eq:X_epsilon_general}
\end{equation}

By construction,
\begin{equation}
    \Omega_\epsilon
    =
    \mathrm{Tr}_{O_M}
    \left[
        \left(
            \tilde P_\epsilon
            \otimes
            \mathcal I_{\mathcal R,O_S}
        \right)
        \left(
            \mathcal I_{\overline S}
            \otimes
            X_\epsilon
        \right)
    \right].
    \label{eq:Omega_recovered_general}
\end{equation}
Therefore the normalization and local-unbiasedness constraints of $X_\epsilon$ follow from those of $\Omega_\epsilon$.

It remains to verify the cone constraint for $X_\epsilon$. Since Eq.~(\ref{eq:X_epsilon_general}) is a congruence transformation, it preserves positive semidefiniteness. For the tight cone $\mathcal C_1$, a separable decomposition of $\Omega_\epsilon$ is mapped to a separable decomposition of $X_\epsilon$, because the congruence acts only on the Hilbert-space factor and maps positive operators to positive operators. Hence
\begin{equation}
    \Omega_\epsilon\in \tilde{\mathcal C}_1
    \quad\Longrightarrow\quad
    X_\epsilon\in\mathcal C_1.
\end{equation}
For $\mathcal C_2$ and $\mathcal C_3$, the same congruence preserves the block relations defining $\mathcal B$ and $\mathcal B^{\prime\prime}$, because it is Hermiticity preserving and acts trivially on the auxiliary space $\mathcal R$. Thus
\begin{equation}
   \Omega_\epsilon\in \tilde{\mathcal C}_k
    \quad\Longrightarrow\quad
    X_\epsilon\in\mathcal C_k,
    \qquad
    k=2,3.
\end{equation}
For the Holevo cone, define
\begin{equation}
    \rho_{\epsilon,\bm\theta}
    =
    N_{\bm\theta}*\tilde P_\epsilon
    =
    \mathrm{Tr}_{\overline S}
    \left[
        \left(
            \tilde P_\epsilon
            \otimes
            \mathcal I_S
        \right)
        \left(
            \mathcal I_{O_M}
            \otimes
            T_{\bm\theta}
        \right)
    \right].
\end{equation}
Using Eq.~(\ref{eq:Omega_recovered_general}), we have
\begin{equation}
    \mathrm{Tr}
    \left[
        X_\epsilon
        \left(
            \Lambda_{ij}
            \otimes
            \rho_{\epsilon,\bm\theta}
        \right)
    \right]
    =
    \mathrm{Tr}
    \left[
        \Omega_\epsilon
        \left(
            \Lambda_{ij}
            \otimes
            T_{\bm\theta}
        \right)
    \right].
\end{equation}
Since $\Omega_\epsilon$ satisfies the Holevo constraint, the right-hand side is zero. Hence
\begin{equation}
    X_\epsilon\in\mathcal C_4.
\end{equation}
Thus $(\tilde P_\epsilon,X_\epsilon)$ is feasible for $Q_k^P$.

The objective values are equal:
\begin{equation}
\begin{aligned}
    \mathrm{Tr}
    \left[
        (W\otimes\rho_{\epsilon,\bm\theta})
        X_\epsilon
    \right]
    &=
    \mathrm{Tr}
    \left[
        (W\otimes T_{\bm\theta})
        \Omega_\epsilon
    \right].
\end{aligned}
\end{equation}
Therefore, for every $\epsilon>0$,
\begin{equation}
    Q_k^P
    \le
    \mathrm{Tr}
    \left[
        (W\otimes T_{\bm\theta})
        \Omega_\epsilon
    \right].
\end{equation}
Taking the limit $\epsilon\to0^+$ gives
\begin{equation}
    Q_k^P
    \le
    \mathrm{Tr}
    \left[
        (W\otimes T_{\bm\theta})
        \Omega^*
    \right].
\end{equation}
Since $\Omega^*$ was an arbitrary feasible solution of $S_k^P$, we obtain
\begin{equation}
    Q_k^P\le S_k^P.
\end{equation}
Together with $S_k^P\le Q_k^P$, this proves
\begin{equation}
    S_k^P=Q_k^P,
    \qquad
    k=1,2,3,4.
\end{equation}

\section{Dual Formulation of Theorem 1}
\label{appendix:proof of theorem 2}

We consider the primal problem defined in Theorem 1 for definite order strategies,
\begin{equation}
\begin{aligned}
\min_{\Omega \in \mathcal{C}_k,\; P\ge 0}\quad
& \mathrm{Tr}\!\big[\Omega (W \otimes T_{\bm{\theta}})\big] \\
\text{s.t.}\quad
& \mathrm{Tr}_{\mathcal R}\!\big[\Omega (|0\rangle\langle 0|\otimes \mathcal{I}_S)\big]
= P\otimes I_{O_S}, \\
& \mathcal A(P)=b,\\
& \frac{1}{2}\mathrm{Tr}\!\Big[
\Omega \big( (|0\rangle\langle i|+|i\rangle\langle 0|)\otimes R_j \big)
\Big]
= \delta_{ij},
\end{aligned}
\label{eq:primal_general_affine}
\end{equation}
where \(\mathcal C_k,~k=2,3,4\) denotes the admissible set of \(\Omega\) defined in Appendix \ref{appendix:Expressions for Cones}, \(\mathcal A(P)=b\) is an affine function corresponding to the constraints for a certain strategy.

For convenience, we define \(R := W \otimes T_{\bm{\theta}},~
H_{ij} := (|0\rangle\langle i|+|i\rangle\langle 0|)\otimes R_j,\) and the linear map $\Phi(\Omega):=
\mathrm{Tr}_{\mathcal R}\!\big[\Omega (|0\rangle\langle 0|\otimes \mathcal{I}_S)\big]$. Then the primal problem can be written compactly as
\begin{equation}
\begin{aligned}
\min_{\Omega \in \mathcal C_k,\; P}\quad
& \mathrm{Tr}(R\Omega) \\
\text{s.t.}\quad
& \Phi(\Omega)-P\otimes I_{O_S}=0, \\
& \mathcal A(P)-b=0, \\
& P \succeq 0, \\
& \frac{1}{2}\mathrm{Tr}(\Omega H_{ij})=\delta_{ij}.
\end{aligned}
\label{eq:primal_general_affine_compact}
\end{equation}

To derive the dual, we introduce a Hermitian operator-valued Lagrange multiplier
\(Y=Y^\dagger\) for the constraint \(\Phi(\Omega)-P\otimes I_{O_S}=0\), a multiplier \(Z\) for the affine constraint \(\mathcal A(P)-b=0\), and real multipliers \(\omega_{ij}\in\mathbb R\) for the scalar constraints \(\frac12\mathrm{Tr}(\Omega H_{ij})=\delta_{ij}\). The Lagrangian is
\begin{equation}
\begin{aligned}
L(\Omega,P;Y,Z,\omega)
={}&\ \mathrm{Tr}(R\Omega)
-\mathrm{Tr}\!\big[Y(\Phi(\Omega)-P\otimes I_{O_S})\big] \\
&-\langle Z,\mathcal A(P)-b\rangle \\
&-\sum_{i,j}\omega_{ij}
\left(
\frac{1}{2}\mathrm{Tr}(\Omega H_{ij})-\delta_{ij}
\right),
\end{aligned}
\label{eq:Lagrangian_general_affine}
\end{equation}
where \(\langle \cdot,\cdot\rangle\) denotes the Hilbert--Schmidt inner product on the codomain of \(\mathcal A\).

Let \(\Phi^*\) and \(\mathcal A^*\) denote the adjoint maps of \(\Phi\) and \(\mathcal A\), respectively, namely,
\begin{equation}
\mathrm{Tr}\!\big[Y\,\Phi(\Omega)\big]
=
\mathrm{Tr}\!\big[\Phi^*(Y)\,\Omega\big]
\qquad
\forall\, \Omega,\;Y,
\end{equation}
and
\begin{equation}
\langle Z,\mathcal A(P)\rangle
=
\mathrm{Tr}\!\big[\mathcal A^*(Z)\,P\big]
\qquad
\forall\, P,\;Z.
\end{equation}
Using these identities, the Lagrangian can be rewritten as
\begin{equation}
\begin{aligned}
L(\Omega,P;Y,Z,\omega)
={}&\ \mathrm{Tr}\!\Bigg[
\Omega\Bigg(
R-\Phi^*(Y)-\frac{1}{2}\sum_{i,j}\omega_{ij}H_{ij}
\Bigg)
\Bigg] \\
&+\mathrm{Tr}\!\big[(\mathrm{Tr}_{O_S}[Y]-\mathcal A^*(Z))P\big] \\
&+\langle Z,b\rangle
+\sum_{i,j}\omega_{ij}\delta_{ij}.
\end{aligned}
\label{eq:Lagrangian_general_affine_rewritten}
\end{equation}

The dual function is obtained by minimizing the Lagrangian over the primal variables:
\begin{equation}
g(Y,Z,\omega)
:=
\inf_{\Omega\in\mathcal C_k,\;P\succeq 0}
L(\Omega,P;Y,Z,\omega).
\end{equation}

Since \(\mathcal C_k\) is a cone, the minimization over \(\Omega\in\mathcal C_k\) is finite if and only if
\begin{equation}
R-\Phi^*(Y)-\frac{1}{2}\sum_{i,j}\omega_{ij}H_{ij}
\in \mathcal C_k^*,
\label{eq:general_affine_dual_cone_C}
\end{equation}
where $\mathcal C_k^*
:=
\{X \mid \mathrm{Tr}(XZ)\ge 0,\ \forall\, Z\in\mathcal C_k\}$ is the dual cone of \(\mathcal C_k\). Similarly, since \(P\succeq 0\), the minimization over \(P\) is finite if and only if
\begin{equation}
\mathrm{Tr}_{O_S}[Y]-\mathcal A^*(Z)\succeq 0.
\label{eq:general_affine_dual_psd}
\end{equation}
Indeed, if \(\mathrm{Tr}_{O_S}[Y]-\mathcal A^*(Z)\not\succeq 0\), then one can choose \(P\succeq 0\) supported on a negative eigenspace of \(\mathrm{Tr}_{O_S}[Y]-\mathcal A^*(Z)\), which makes
\(\mathrm{Tr}[(\mathrm{Tr}_{O_S}[Y]-\mathcal A^*(Z))P]\) arbitrarily negative. Conversely, if \(\mathrm{Tr}_{O_S}[Y]-\mathcal A^*(Z)\succeq 0\), then
\(\mathrm{Tr}[(\mathrm{Tr}_{O_S}[Y]-\mathcal A^*(Z))P]\ge 0\) for all \(P\succeq 0\), and the infimum is attained at \(P=0\), giving value \(0\).

Therefore,
\begin{equation}
g(Y,Z,\omega)=
\begin{cases}
\displaystyle
\langle Z,b\rangle+\sum_{i,j}\omega_{ij}\delta_{ij},
& \text{if \eqref{eq:general_affine_dual_cone_C} and \eqref{eq:general_affine_dual_psd} hold}, \\[1.2ex]
-\infty,
& \text{otherwise}.
\end{cases}
\end{equation}

We thus arrive at the Lagrange dual problem
\begin{equation}
\begin{aligned}
\max_{Y=Y^\dagger,\; Z,\; \omega_{ij}\in\mathbb R}\quad
& \langle Z,b\rangle+\sum_{i,j}\omega_{ij}\delta_{ij} \\
\text{s.t.}\quad
& W\otimes T_{\bm\theta}
-\Phi^*(Y)
-\frac{1}{2}\sum_{i,j}\omega_{ij}
\big((|0\rangle\langle i|+|i\rangle\langle 0|)\otimes R_j\big)
\in \mathcal C_k^*, \\
& \mathrm{Tr}_{O_S}[Y]-\mathcal A^*(Z)\succeq 0.
\end{aligned}
\label{eq:dual_general_affine}
\end{equation}

This is the exact dual formulation of the primal problem in Eq.~\eqref{eq:primal_general_affine}. Precisely, $\Phi^*(Y)$ can be chosen as
\begin{equation}
    \Phi^*(Y)=\ket{0}\bra{0}_{\mathcal{R}}\otimes Y.
\end{equation}

\subsection{Sequential Strategies}
We now specialize the affine constraint to the recursive $N$-step form. Let the strategy comb and sub-comb be
\begin{equation}
P^{(N)},\; P^{(N-1)},\; \dots,\; P^{(1)},
\end{equation}
with constraints
\begin{equation}
P = P^{(N)},
\qquad
\text{Tr}_{2k-1}[P^{(k)}]=\mathcal{I}_{2k-2}\otimes P^{(k-1)}
\quad (k=2,\dots,N),
\qquad
\mathrm{Tr}[P^{(1)}]=1,
\label{eq:recursive_constraints_N}
\end{equation}
together with $P \succeq 0$, $P^{(k)} \succeq 0
\quad (k=1,\dots,N)$.

The primal problem therefore becomes
\begin{equation}
\begin{aligned}
\min_{\Omega \in \mathcal C_k,\, P,\, P_N,\dots,P_1}\quad
& \mathrm{Tr}(R\Omega) \\
\text{s.t.}\quad
& \Phi(\Omega)-P^{(N)}\otimes \mathcal{I}_{2N}=0, \\
& \text{Tr}_{2k-1}[P^{(k)}]-\mathcal{I}_{2k-2}\otimes P^{(k-1)}=0,
\qquad k=2,\dots,N, \\
& \mathrm{Tr}[P^{(1)}]-1=0, \\
& P_k\succeq0\ \ (k=1,\dots,N), \\
& \frac12\mathrm{Tr}(\Omega H_{ij})=\delta_{ij},\qquad i,j=1,2,\dots,d.
\end{aligned}
\label{eq:primal_recursive_N}
\end{equation}

Thus the dual problem is given by
\begin{equation}
\begin{aligned}
\max_{Q^{(N)},\dots,Q^{(1)},\lambda,\omega}\quad
& \lambda+\sum_{i,j}\omega_{ij}\delta_{ij} \\
\text{s.t.}\quad
& W\otimes T_{\bm\theta}
-|0\rangle\langle0|_{\mathcal{R}}\otimes Q^{(N)}
-\frac12\sum_{i,j}\omega_{ij}
\big((|0\rangle\langle i|_{\mathcal{R}}+|i\rangle\langle0|_{\mathcal{R}})\otimes R_j\big)
\in\mathcal C_k^*, \\
& \mathrm{Tr}_{2k}[Q^{(k)}]-Q^{(k-1)}\otimes \mathcal{I}_{2k-1}\succeq0,
\qquad k=2,\dots,N, \\
& \mathrm{Tr}_{2}[Q^{(1)}]-\lambda \mathcal{I}_1\succeq0, \\
& Q^{(k)}=Q^{(k)\dagger}\ \ (k=1,\dots,N),
\quad \lambda\in\mathbb R,
\quad \omega_{ij}\in\mathbb R,
\end{aligned}
\label{eq:dual_recursive_N}
\end{equation}
where $Y\in \mathcal H_{1,2,\dots,2N}$ and $Q^{(k)}$ is defined in $\mathcal H_{1,2,\dots,2k-1,2k}$ for $k=1,2,\dots, N$.

\subsection{Parallel Strategies}
For parallel strategies, which can be formulated as $1$-step sequential strategies with
\begin{equation}
P = P^{(1)},
\qquad
\mathrm{Tr}[P^{(1)}]=1,
\label{eq:recursive_constraints_1}
\end{equation}
with $P^{(1)} \succeq 0$ and $P^{(1)}\in \mathcal{H}_{1,3,\dots,2N-1}$.

The dual problem is given by
\begin{equation}
\begin{aligned}
\max_{Q,\lambda,\omega}\quad
& \lambda+\sum_{i,j}\omega_{ij}\delta_{ij} \\
\text{s.t.}\quad
& W\otimes T_{\bm\theta}
-|0\rangle\langle0|_{\mathcal{R}}\otimes Q
-\frac12\sum_{i,j}\omega_{ij}
\big((|0\rangle\langle i|_{\mathcal{R}}+|i\rangle\langle0|_{\mathcal{R}})\otimes R_j\big)
\in\mathcal C_k^*, \\
& \mathrm{Tr}_{2,4,\dots,2N}[Q]-\lambda I_{1,3,\dots,2N-1}\succeq0, \\
& Q=Q^\dagger\ \ (k=1,\dots,N),
\quad \lambda\in\mathbb R,
\quad \omega_{ij}\in\mathbb R.
\end{aligned}
\label{eq:dual_recursive_1}
\end{equation}
where $Q\in \mathcal H_{1,2,\dots,2N}$.

\section{Numerical Verification of Optimal Sequential and Parallel Strategies for multi-dimensional quantum magnetometry under unitary evolution}
\label{appendix:NV for seq and par}

\subsection{Optimal Control of Sequential Feedback Scheme}

Following Ref.~\cite{yuan2016sequential}, the optimal control in the sequential feedback scheme is given by
\begin{equation}
    U_1=U_2=\dots=U_N=U_{\mathcal{M}}^\dagger(\hat{\bm{\theta}},t),
\end{equation}
where $U_{\mathcal{M}}(\hat{x},t)=e^{-iH(\bm{\theta})t}\otimes \mathcal{I}_{\mathcal{M}}$. The corresponding probe state is the maximally entangled state $\ket{\Phi_0}=\frac{\ket{00}+\ket{11}}{\sqrt{2}}$. We numerically verify this analytical structure by constructing a sequence of isometries using the minimal-ancilla realization of quantum combs developed in \cite[Theorem~1,~2]{bisio2011minimal}. For an $N$-step quantum process $\mathscr{P}$ with CJ operator $P$ defined on
$\mathsf{Comb}[
(\emptyset,\mathcal{H}_1),
(\mathcal{H}_2,\mathcal{H}_3),
\dots,(\mathcal{H}_{2N-2},\mathcal{H}_{2N-1})]$, there exist isometries $\{V^{(k)}\}_{k=1}^N$ such that
\begin{equation}
    \mathscr{P}(\rho)=\mathrm{Tr}_{\mathcal{M}_{N}}\left[\left(V^{(N)}\otimes \mathcal{I}_{2,4,\dots,2N-2}\right)\cdots\left(V^{(1)}\otimes \mathcal{I}_{3,5,\dots,2N-1}\right)\rho \left(V^{(1)}\otimes \mathcal{I}_{3,5,\dots,2N-1}\right)^\dagger \cdots \left(V^{(N)}\otimes \mathcal{I}_{2,4,\dots,2N-2}\right)^\dagger\right]
\end{equation}
for an arbitrary input $\rho \in \mathcal{L}(\bigotimes_{i=1}^N\mathcal{H}_{2i-1})$. The isometries $V^{(k)}$, acting from $\mathcal{H}_{2k-2}\otimes\mathcal{M}_{k-1}$ to $\mathcal{H}_{2k-1}\otimes\mathcal{M}_{k}$, are obtained from the sub-combs $P^{(k)}$ as
\begin{equation}
\begin{aligned}
    V^{(k)} = \left\{\mathcal{I}_{2k-1}\otimes [P^{(k)\frac{1}{2}}]^*[P^{(k-1)-\frac{1}{2}}]^*\right\}\left(|I\rrangle_{(2k-1)(2k-1)^\prime}T_{(2k-2)\to (2k-2)^\prime}\right),
\end{aligned}
\end{equation}
where the minimal memory space is chosen as $\mathcal{M}_{k} = \mathsf{Supp}(P^{(k)*})$, $|I\rrangle_{n,m}=\sum_i \ket{i}_{n}\ket{i}_{m}$, and $T_{n\to m}=\sum_j\ket{j}_{m}\bra{j}_{n}$. We also set $\mathcal{H}_0=\mathbb{C}$ and $\mathcal{M}_0=\mathbb{C}$. Under this convention, $V^{(1)}$ prepares the probe state, while $V^{(k)}$, $k=2,\dots,N$, implements the intermediate controls. From the explicit construction, the minimal dimension of the memory space required to implement the sequential strategy $P=P^{(N)}$ is
\begin{equation}
\mathsf{dim}(\mathcal{H}_{\mathcal{M}_N} ) = 
\mathsf{rank}(P^{(N)})\le \mathsf{dim}(P^{(N)})=\Pi_{i=1}^{2N-1}\mathsf{dim}(\mathcal{H}_i).
\end{equation}

For the case $N = 2$ discussed in the main text, the minimal memory dimensions are $\mathsf{dim}(\mathcal{M}_1)=2$ and $\mathsf{dim}(\mathcal{M}_2)=8$. We label $\mathcal{M}_2=\mathcal{H}_{3^\prime}\otimes\mathcal{H}_{2^\prime}\otimes\mathcal{H}_{1^\prime}$. The corresponding isometries are
\begin{equation}
\begin{aligned}
    V^{(1)} 
    =\sum_i \ket{i}_1\otimes [P^{(1)\frac{1}{2}}]^*\ket{i}_{\mathcal{M}_1},
\end{aligned}
\end{equation}
and
\begin{equation}
\label{V2}
\begin{aligned}
    V^{(2)} 
    =\sum_{i,j} \ket{i}_3\bra{j}_2\otimes \left\{[P^{(2)\frac{1}{2}}]^*\left(\ket{i}_{3^\prime}\otimes \ket{j}_{2^\prime}\otimes [P^{(1)-\frac{1}{2}}]^*\right)\right\}.  
\end{aligned}
\end{equation}

The analytical optimal feedback control implies $P^{(1)}=\frac{1}{2}I$. Moreover, $V^{(2)}$ is identified with the inverse operation of $e^{-iH(\bm{\theta})t}\otimes\mathcal{I}_{\mathcal{M}}$, namely
$V^{(2)}=\mathcal{E}^\dagger \otimes \ket{0}_{3^\prime}\otimes \ket{0}_{2^\prime} \otimes \mathcal{I}_{1^\prime}$. The evolution on the ancillary space is chosen to be the identity, since it does not affect the final precision bound. Substituting this expression into Eq.~(\ref{V2}), we obtain the square-root matrix of $P^{(2)}$,
\begin{equation}
    P^{(2)\frac{1}{2}}=\frac{1}{\sqrt{2}}\left( \sum_{i,j}\sum_{i^\prime,j^\prime} e_{i,j}^*e_{i^\prime,j^\prime}~ \ket{i^\prime,j^\prime}_{3^\prime2^\prime}\bra{i,j}_{3^\prime2^\prime} \right )\otimes \mathcal{I}_{1^\prime},
\end{equation}
where $e_{i,j}=\bra{j}_2\mathcal{E}\ket{i}_{3}$. Since the operator in the parentheses is rank one, we obtain the analytic feedback-control construction of $P^{(2)}$:
\begin{equation}
\label{P2}
    P^{(2)}_{\text{FB}}=\frac{1}{2}\left( \sum_{i,j}\sum_{i^\prime,j^\prime} e_{i,j}^*e_{i^\prime,j^\prime}~ \ket{i^\prime,j^\prime}_{3^\prime2^\prime}\bra{i,j}_{3^\prime2^\prime} \right )\otimes \mathcal{I}_{1^\prime}.
\end{equation}

For the noiseless case with \(W=I\), \(\alpha=2.5\), \(\theta=1.3\), and \(\phi=\pi/4\), the sub-combs obtained from the SDP agree with the analytical optimal feedback-control construction. In particular, the optimization gives \(P^{(1)}_{\text{SDP}}=\frac{1}{2}I\). Moreover, the resulting analytical \(P^{(2)}_{\text{FB}}\) matches the optimized \(P^{(2)}_{\text{SDP}}\) returned by the optimization within numerical precision. The solution is displayed explicitly as

\begin{equation}
   \label{P2_value}
    P^{(2)}_{\text{FB}}=P^{(2)}_{\text{SDP}}=\begin{bmatrix}
 0.334 & -0.169-0.164\mathrm{i} & 0.219-0.086\mathrm{i} & 0.308+0.128\mathrm{i} \\
 -0.169+0.164\mathrm{i} & 0.166 & -0.069+0.151\mathrm{i} & -0.219+0.086\mathrm{i} \\
 0.219+0.086\mathrm{i} & -0.069-0.151\mathrm{i} & 0.166 & 0.169+0.164\mathrm{i} \\
 0.308-0.128\mathrm{i} & -0.219-0.086\mathrm{i} & 0.169-0.164\mathrm{i} & 0.334
\end{bmatrix}\otimes \begin{bmatrix}
    1&0\\0&1
    \end{bmatrix}
\end{equation}

\subsection{Optimal Probe State of Parallel Strategy}

\begin{figure}[h]
   \centering
    \captionsetup{justification=raggedright,singlelinecheck=false}
    \includegraphics[width=0.25\linewidth]{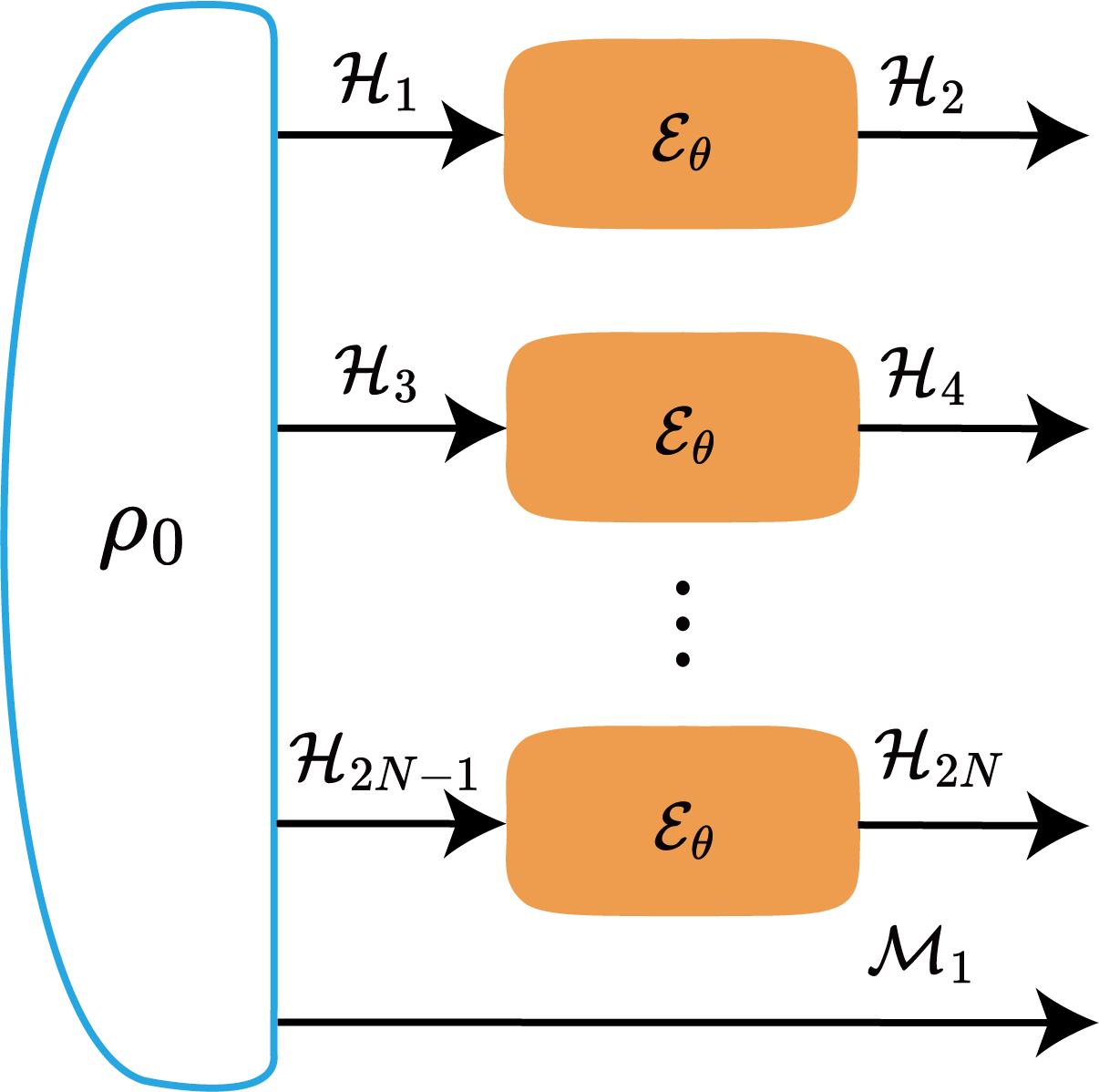}
    \caption{\textbf{Parallel scheme with $N$ quantum channels.} $\tilde{P}=\rho_0\in \mathcal{M}_1\bigotimes_{i=1}^{N}\mathcal{H}_{2i-1}$ is the probe state thus the parallel strategy corresponds to the state preparation stage of the metrological process. The reduced state $\rho^{(1,3,\dots,2N-1)}$ on the input system $\bigotimes_{i=1}^{N}\mathcal{H}_{2i-1}$ corresponds to the sub-comb $P^{(1)}=\text{Tr}_{\mathcal{M_1}}[\rho_0]$.}
    \label{Appendix-figure-parallel}
\end{figure}

For the parallel scheme, the optimization returns the reduced state of the optimal probe on the input systems $\bigotimes_{i=1}^{N}\mathcal{H}_{2i-1}$. We denote this reduced state by
$P^{(1)}=\rho_{1,3,\dots,2N-1}=\mathrm{Tr}_{\mathcal{M}1}[\rho_0]$, where the full optimal probe state satisfies
$\rho_0\in \mathcal{M}_1\bigotimes_{i=1}^{N}\mathcal{H}_{2i-1}$, as illustrated in Fig.~\ref{Appendix-figure-parallel}.

According to Ref.~\cite{hou2020minimal}, for the two-channel case the optimal reduced two-qubit state on $\mathcal{H}_{1,3}$ is given by
\begin{equation}
\label{analytical_par_state}
\begin{aligned}
\rho^{(1,3)}=\frac{1}{4}\left[I^{(j,k)}+\tilde{r}_{\alpha \alpha} \sigma_{\alpha}^{(1)} \otimes \sigma_{\alpha}^{(3)}+\tilde{r}_{\theta \theta} \sigma_{\theta}^{(1)} \otimes \sigma_{\theta}^{(3)}+\tilde{r}_{\phi \phi} \sigma_{\phi}^{(1)} \otimes \sigma_{\phi}^{(3)}\right],
\end{aligned}
\end{equation}
where the coefficients are
\begin{equation}
\begin{array}{l}
\tilde{r}_{\alpha \alpha}=\frac{(N+1) \sqrt{w_{\alpha}}-\frac{\sqrt{w_{\theta}}}{|\sin \alpha|}-\frac{\sqrt{w_{\phi}}}{|\sin \alpha \sin \theta|}}{(N-1)\left(\sqrt{w_{\alpha}}+\frac{\sqrt{w_{\theta}}}{|\sin \alpha|}+\frac{\sqrt{w_{\phi}}}{|\sin \alpha \sin \theta|}\right)}, \
\tilde{r}_{\theta \theta}=\frac{(N+1) \frac{\sqrt{w_{\theta}}}{|\sin \alpha|}-\sqrt{w_{\alpha}}-\frac{\sqrt{w_{\phi}}}{|\sin \alpha \sin \theta|}}{(N-1)\left(\sqrt{w_{\alpha}}+\frac{\sqrt{w_{\theta}}}{|\sin \alpha|}+\frac{\sqrt{w_{\phi}}}{|\sin \alpha \sin \theta|}\right)}, \
\tilde{r}_{\phi \phi}=\frac{(N+1) \frac{\sqrt{w_{\phi}}}{|\sin \alpha \sin \theta|}-\sqrt{w_{\alpha}}-\frac{\sqrt{w_{\theta}}}{|\sin \alpha|}}{(N-1)\left(\sqrt{w_{\alpha}}+\frac{\sqrt{w_{\theta}}}{|\sin \alpha|}+\frac{\sqrt{w_{\phi}}}{|\sin \alpha \sin \theta|}\right)},
\end{array}
\end{equation}
and $\sigma_{i}^{(j)}=\bm{n}_{i}\cdot\bm{\sigma}^{(j)}$, with $\bm{\sigma}^{(j)}=(\sigma_x^{(j)},\sigma_y^{(j)},\sigma_z^{(j)})$ denoting the Pauli operators acting on $\mathcal{H}_j$. The vectors $\bm{n}_{i}$ are defined as
\begin{equation}
\begin{array}{l}
\bm{n}_\alpha=(\sin\theta\cos\phi,\sin\theta\sin\phi,\cos\theta),\
\bm{n}_\theta=\cos\alpha \bm{n}_1-\sin\alpha \bm{n}_2,\
\bm{n}\phi=\cos\alpha \bm{n}_2+\sin\alpha \bm{n}_1,
\end{array}
\end{equation}
where
\begin{equation}
\begin{array}{l}
\bm{n}_1=(\cos\theta\cos\phi,\cos\theta\sin\phi,-\sin\theta),\
\bm{n}_2=(-\sin\phi,\cos\phi,0).
\end{array}
\end{equation}

For the noiseless case the reduced state obtained from our SDP agrees with the analytical expression in Eq.~(\ref{analytical_par_state}). For instance, when $W=I$, $\alpha=2.5$, $\theta=1.3$, and $\phi=\pi/4$, the numerically obtained optimal reduced state is
\begin{equation}
\label{P1_value}
    P^{(1)}=\rho^{(1,3)}=\begin{bmatrix}
        0.373 & -0.028+0.031 i & -0.028+0.031 i & 0.066+0.135 i \\
-0.028-0.031 i & 0.127 & 0.127 & 0.028-0.031 i \\
-0.028-0.031 i & 0.127 & 0.127 & 0.028-0.031 i \\
0.066-0.135 i & 0.028+0.031 i & 0.028+0.031 i & 0.373
    \end{bmatrix},
\end{equation}    
which agrees with the analytical expression in Eq.~(\ref{analytical_par_state}). This confirms that our SDP formulation faithfully recovers the same optimal reduced probe state as the analytical approach. Any purification of $\rho^{(1,3)}$ on $\mathcal{H}_{1,3,\mathcal{M}_1}$ can then serve as an optimal initial state.

Importantly, for the parameter setting considered here, the analytical coefficient
\(\tilde r_{\alpha\alpha}=-0.0919\). Consequently, the explicit qutrit-ancilla GHZ-type construction proposed in Ref.~\cite{hou2020minimal}, for which the correlation coefficients are represented as
\(\tilde r_{xx}=\lvert s_x\rvert^2\), cannot realize the optimal correlations in this regime. This does not imply that the state in Eq.~(\ref{analytical_par_state}) is unphysical, since the quantities
\(\tilde r_{xx}\) are correlation coefficients rather than probabilities. Indeed, the full density matrix in Eq.~(\ref{P1_value}) is positive semidefinite and is directly recovered by our optimization.

Our result therefore shows that the non-negativity condition required by the explicit probe ansatz of Ref.~\cite{hou2020minimal} is sufficient, but not necessary, for finite-\(N\) attainability of the analytical bound. For the present \(N=2\) example, the SDP recovers both the analytical precision benchmark and an explicit physical probe realizing the corresponding optimal correlation structure. This observation is consistent with the recent analysis of Ref.~\cite{zhou2026unifiedcomputableapproachoptimal}, which shows that the restricted heuristic probe states considered previously may be suboptimal for small \(N\), while the analytical parallel lower bound can nevertheless remain tight.

\end{document}